\documentclass[prb,twocolumn,amsmath,amssymb,floatfix,superscriptaddress]{revtex4-2}
\usepackage{graphicx} 
\usepackage{booktabs}
\usepackage{color}
\usepackage[version=4]{mhchem}
\usepackage{xspace, bm, braket}
\usepackage{siunitx}
\usepackage{verbatim}
\usepackage{subfigure}
\usepackage[colorlinks, citecolor=blue, linkcolor=blue, urlcolor=blue, breaklinks]{hyperref}

\bibpunct{[}{]}{,}{n}{}{}
\newcommand{\etal}{\textit{et al.}}
\newcommand{\angstrom}{\textup{\AA}}

\usepackage[svgnames]{xcolor}
\newif\ifshowcomments\showcommentstrue
\def\beq{\begin{equation}}
\def\eeq{\end{equation}}

\begin{document}
	\title{Oxygen vacancies at the origin of pinned moments in oxide interfaces: the example of tetragonal CuO/SrTiO$_3$}
		\author{Benjamin Bacq-Labreuil}
		\email{benjamin.bacq-labreuil@polytechnique.edu}
		\affiliation{CPHT,  CNRS,  Ecole  Polytechnique,  Institut  Polytechnique  de  Paris,  F-91128  Palaiseau,  France}
		\author{Benjamin Lenz}
		\email{benjamin.lenz@sorbonne-universite.fr}
		\affiliation{Institut de Min{\'e}ralogie, de Physique des Mat{\'e}riaux et de Cosmochimie, Sorbonne Universit{\'e}, CNRS, MNHN, IRD, 4 Place Jussieu, 75252 Paris, France}
		\author{Silke Biermann}
		\email{silke.biermann@polytechnique.edu}
		\affiliation{CPHT,  CNRS,  Ecole  Polytechnique,  Institut  Polytechnique  de  Paris,  F-91128  Palaiseau,  France}
		\affiliation{Coll{\`e}ge  de  France,  11  place  Marcelin  Berthelot,  75005  Paris,  France}
		\affiliation{Department  of  Physics,  Division  of  Mathematical  Physics,Lund  University,  Professorsgatan  1,  22363  Lund,  Sweden}
		\affiliation{European  Theoretical  Spectroscopy  Facility,  Europe}
	\date{\today}
	
\begin{abstract}
	Obtaining an accurate theoretical description of the emergent phenomena in oxide heterostructures is a major challenge.  
	Recently, intriguing paramagnetic spin and pinned orbital moments have been discovered by x-ray magnetic circular dichroïsm measurements at the Cu $L_{2,3}$-edge of a tetragonal CuO/SrTiO$_3$ heterostructure. 
	Using first principles calculations, we propose a scenario that explains both types of moments, based on the formation of oxygen vacancies in the TiO$_2$ interface layer. 
	We show the emergence of a paramagnetic 2D electron gas hosted in the interface CuO layer.
	It is invisible at the Ti $L_{2,3}$-edge since the valence of the Ti atoms remains unchanged.
	Strong structural distortions breaking both the local and global fourfold rotation $C_4$ symmetries at the interface lead to the in-plane pinning of the Cu orbital moment close to the vacancy.
	Our results, and in particular the pinning of the orbital moment, may have implications for other systems, especially monoxide/dioxide interfaces with similar metal-oxygen bond length and weak spin-orbit coupling.

\end{abstract}
	
	\maketitle

\section{Introduction}

Oxide heterostructures are promising hosts for a large range of applications~\cite{hwang2012, kumah2020}, and offer countless possible configurations allowed by epitaxial thin-film deposition techniques~\cite{koster2015}.
A particular example are copper monoxide CuO thin films, which grow in a tetragonal crystal structure on a SrTiO$_3$ (STO) substrate~\cite{siemons2009,zhong2021}, whereas its bulk counterpart displays a low-symmetry monoclinic structure~\cite{asbrink1970}.
Tetragonal CuO (t-CuO) is made of a staggered stacking of 2D CuO planes along the $c$ axis.
Its tetragonal distortion is characterized by the apical Cu-O distance being 1.37 times larger than the basal distance~\cite{siemons2009, samal2014}.
t-CuO is arguably the cuprate with the simplest crystal structure, making it an ideal candidate for connecting the low-energy effective models for cuprates~\cite{anderson1987, zaanen1985, emery1987, zhang1988, andersen1995, hirayama2018b} to the real materials. 

\begin{figure*}
	\includegraphics[width=\linewidth]{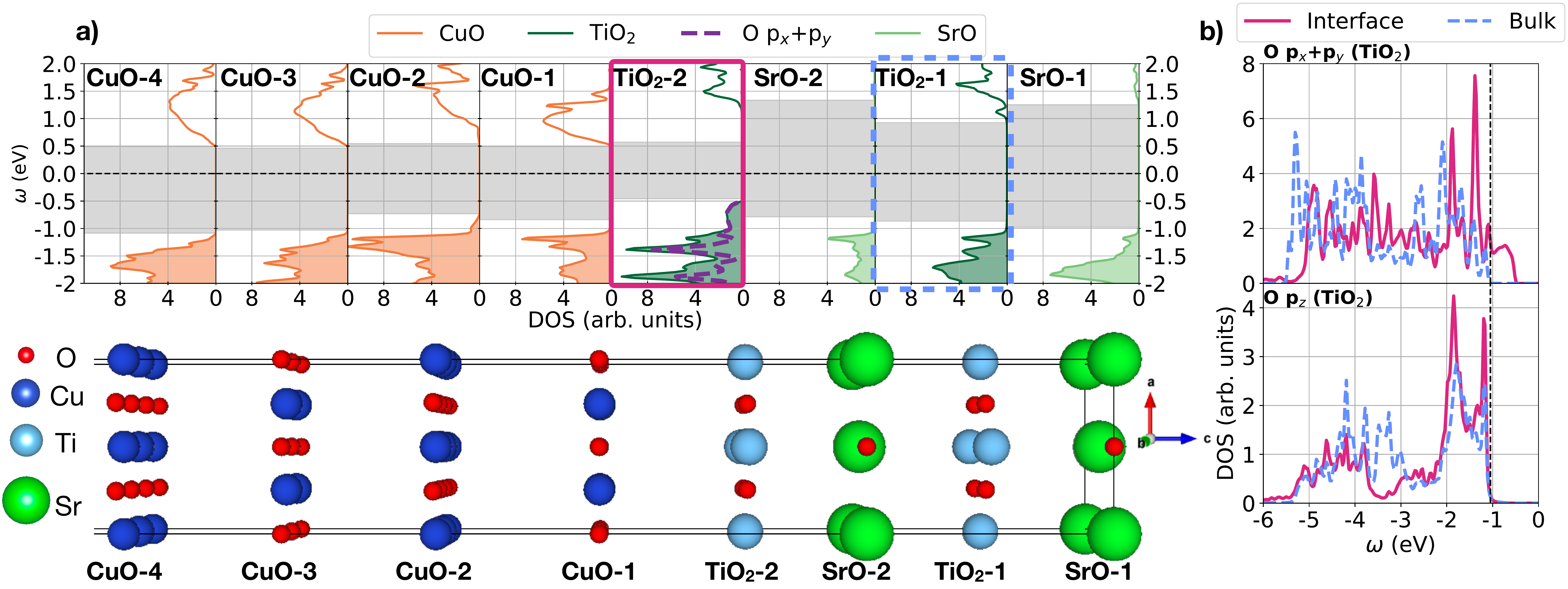}
	\caption{(a) Layer-resolved DOS calculated for the t-CuO/STO junction without defect. The colored shaded area depicts the occupied states, and the gray shaded patches highlight the gap in each layer. (b) Projected DOS on the $p_{x,y,z}$ orbitals of the O atoms in the two TiO$_2$ layers corresponding to the two framed panels in (a).}
	\label{fig:bare_interface}
\end{figure*}
A key step in this direction is the understanding of the properties of undoped t-CuO.
The experimentally observed tetragonal distortions were reproduced using density functional theory (DFT) with hybrid functionals~\cite{chen2009}, and traced back to Jahn-Teller orbital ordering~\cite{peralta2009,wang2014}.
In agreement with resonant inelastic x-ray scattering (RIXS)~\cite{moser2015} and muon spin resonance measurements~\cite{hernandez2021}, it was shown that the CuO layers display an antiferromagnetic stripe order~\cite{peralta2009,chen2009,wang2014}.
A first application of strongly-correlated electron techniques to t-CuO~\cite{bramberger2022} gave an explanation to the sublattice decoupling observed in RIXS~\cite{moser2015}, angle-resolved photoemission spectroscopy~\cite{moser2014}, as well as scanning tunnelling microscopy (STM)~\cite{zhong2021}. 

This set of experimental and theoretical works would yield a coherent understanding of undoped t-CuO, if it was not for the recent finding of an isotropic paramagnetic spin and pinned orbital moment in t-CuO/STO samples, by X-ray magnetic circular dichroïsm (XMCD) measurements at the Cu $L_{2,3}$-edge~\cite{hernandez2021}. 
The existence of these moments is puzzling since t-CuO is antiferromagnetically ordered and does not contain any element with strong spin-orbit coupling (SOC).
Hernandez~$\etal$ has advanced a scenario in Ref.~\onlinecite{hernandez2021}, in which t-CuO would be composed of ferromagnetically ordered CuO layers, which are antiferromagnetically stacked along the $c$ axis (out-of-plane). 
The last CuO layer would be paramagnetic hence would follow the magnetic field, while another layer would be uncompensated and therefore would yield the pinned moments. 
Although this scenario would qualitatively account for the experimental observations, it encounters several difficulties: (i)~the ferromagnetic ordering inside each layer is in contradiction with all previous experimental~\cite{moser2015} and theoretical~\cite{peralta2009,chen2009,wang2014,bramberger2022} findings, (ii)~the uncompensated layer is composed of spin moments and not orbital ones, whereas the observed pinned moment is mainly of orbital character, (iii)~there is no reason why one layer should have a paramagnetic behavior, especially if it is ferromagnetically ordered. 
An alternative explanation would reside in the existence of a 2D electron gas (2DEG) at the t-CuO/STO interface, as was found in other heterostructures such as BiMnO$_3$/STO, LaAlO$_3$/STO and $\gamma$-Al$_2$O$_3$/STO~\cite{salluzzo2013,lee2013,mardegan2019}.
However, in these studies the spin moment is observed at the Ti $L_{2,3}$-edge, corresponding to the existence of a Ti$^{3+}$ valence configuration, contrary to t-CuO/STO for which no Ti$^{3+}$ signal was found~\cite{hernandez2021}. 
The 2DEG scenario was therefore ruled out in Ref.~\onlinecite{hernandez2021}, leaving the question of the origin of these moments open. 

\begin{figure}
	\includegraphics[width=\linewidth]{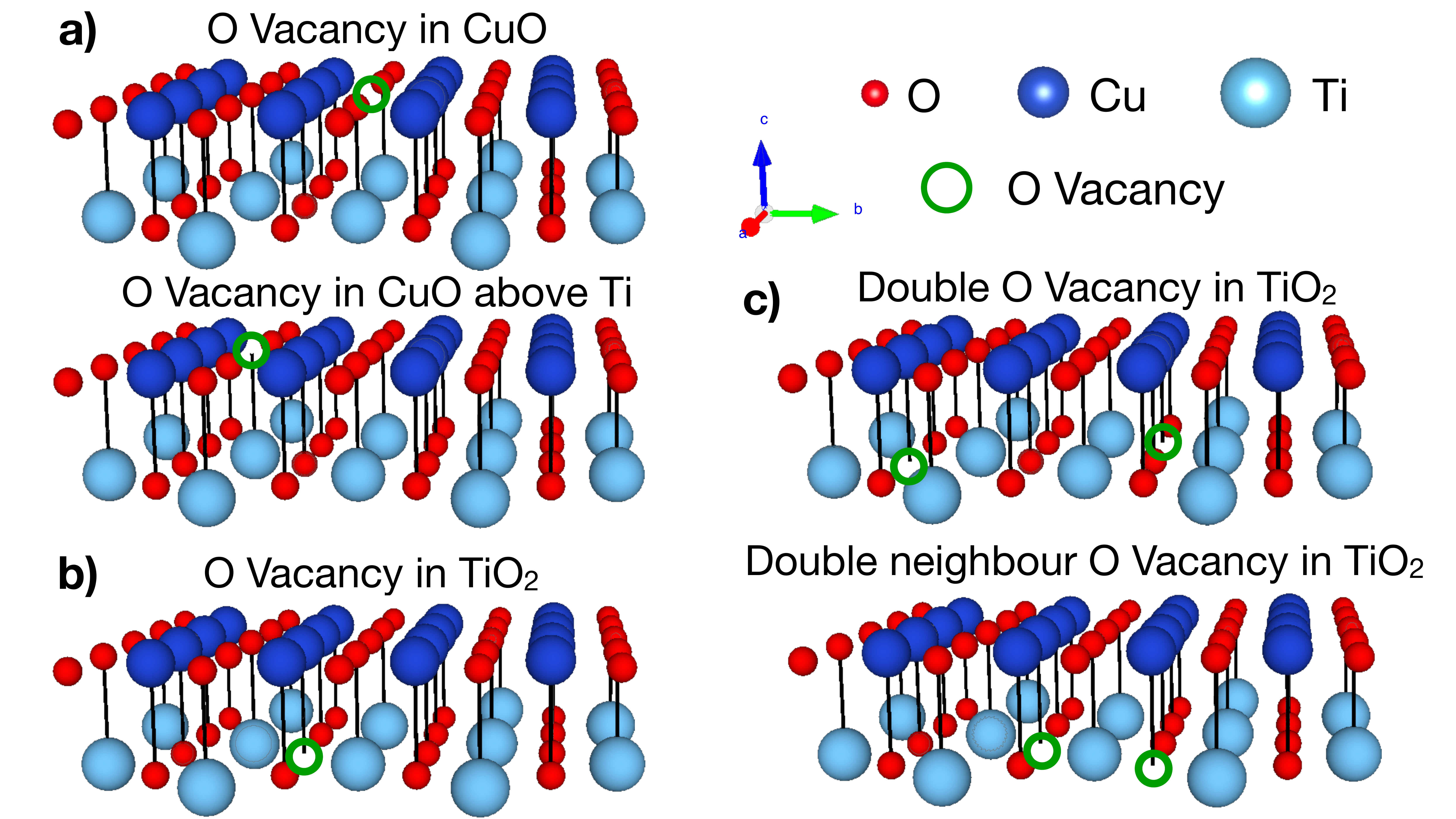}
	\caption{(a)-(c) Zoom on the two interface CuO-1 and TiO$_2$-2 layers presenting the 5 vacancy configurations considered in this work. The black lines between O and Cu/Ti atom show that half of the O in the CuO-1 layer do not have a Ti neighbor.}
	\label{fig:vacancy_struct}
\end{figure} 
In this paper, using first principles DFT$+U$ calculations for the t-CuO/STO heterostructure, we show that the 2DEG scenario is in fact \emph{not incompatible} with the recent XMCD measurements. 
Since oxygen vacancies can generate a 2DEG, as was seen experimentally for AlO$_x$/ZnO~\cite{rodel2018}, STO~\cite{santander2011} and $\gamma$-Al$_2$O$_3$/STO~\cite{chen2013a}, as well as magnetic moments at the surface of STO ~\cite{taniuchi2016,altmeyer2016}, we study the formation of oxygen vacancies at the interface. 
By comparing different vacancy configurations, we show that a spin polarized 2DEG hosted in the CuO layer can emerge when the defect is inserted in the TiO$_2$ interface layer. 
Most importantly, in such a case the overall valence of the Ti atoms remains unchanged with respect to the bulk, such that the 2DEG would be invisible at the Ti $L_{2,3}$-edge.
If in our calculations the 2DEG is spin polarized due to the finite size of our unit cells and the specific arrangement of stripes and vacancies, it would be truly paramagnetic in the real material since the vacancies are randomly distributed. 
Moreover, large distortions are induced in the interface CuO layer,
leading to the breaking of both the local and global $C_4$ symmetries in the CuO layer above the vacancy.
This translates into a pinning of the in-plane component of the orbital moment.
Our scenario therefore provides a mechanism for the paramagnetic spin moment, being a consequence of the emergence of a 2DEG, as well as for the pinned orbital moment, resulting from the breaking of the local and global $C_4$ symmetries.

\begin{figure*}
	\includegraphics[width=\linewidth]{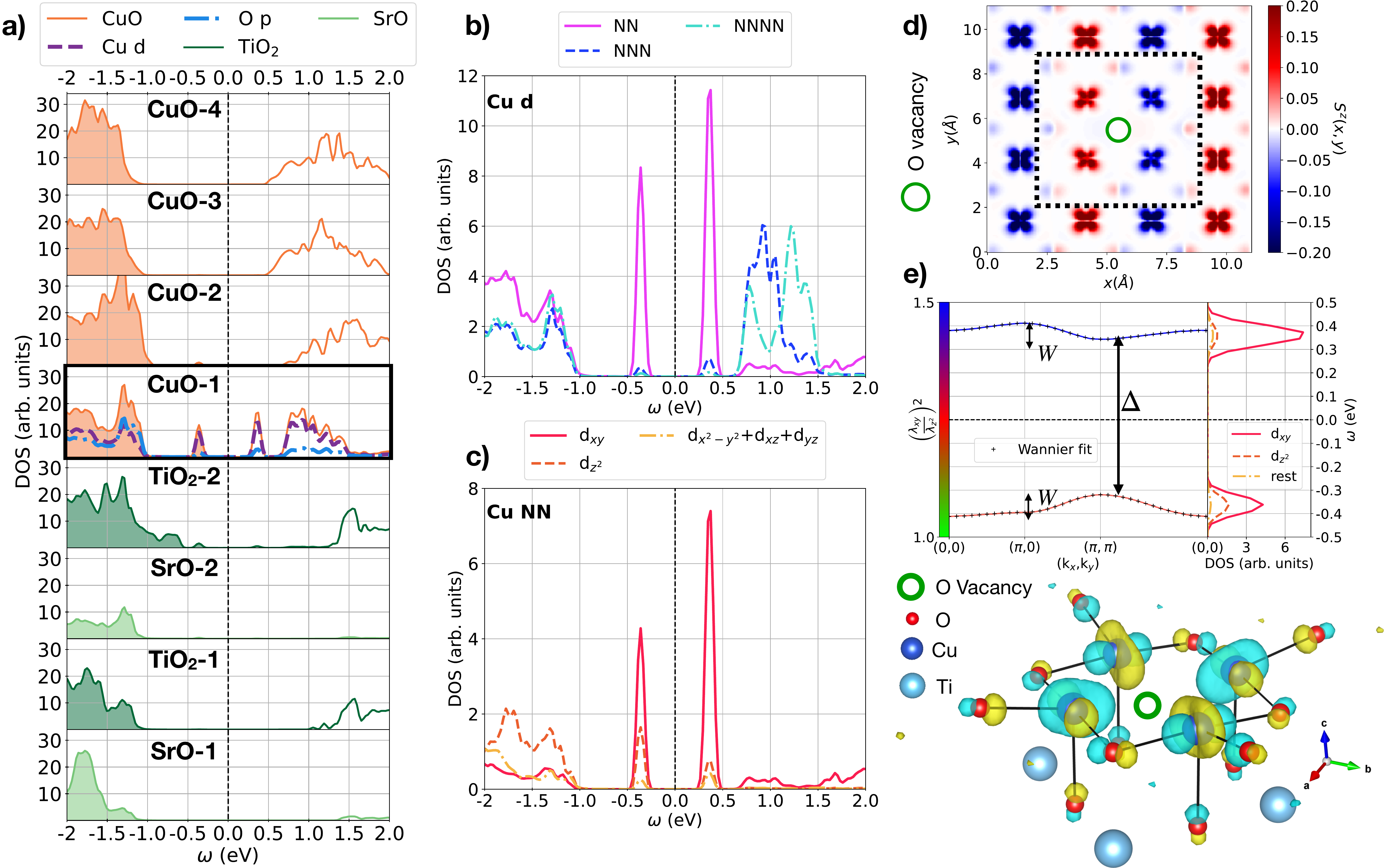}
	\caption{(a) Layer-resolved DOS calculated for the case of an O vacancy in the CuO-1 layer. (b)-(c) Site and orbital projected DOS for the Cu atoms at the CuO-1 interface layer corresponding to the framed panel in (a). (d) Spin density map of the interface CuO-1 layer. (e) Band structure and projected DOS zoomed on the in-gap states along with the fitted Wannier bands (\emph{top}) resulting in the Wannier orbitals showed in the \emph{bottom} around the vacancy, which are centered on the four NN Cu sites highlighted with the dotted frame in (d). The band color depicts the ratio $\left(\frac{\lambda_{xy}}{\lambda_{z^2}}\right)^2$ between the contributions from $d_{xy}$ and $d_{z^2}$.}
	\label{fig:vac_CuO}
\end{figure*}
\section{Model and Method}
To simulate the t-CuO/STO junction we use unit cells of the type shown in Fig.~\ref{fig:bare_interface}(a): the CuO layers are stacked onto a 2 unit cell-thick TiO$_2$-terminated STO substrate~\cite{franchini2011,drera2019,samal2014}.
We apply DFT$+U$~\cite{liechtenstein1995} using the Perdew–Burke–Ernzerhof (PBE) functional~\cite{perdew1996, *perdew1997}, with a local Hubbard interaction $U =6 \si{eV}$ on the Cu $d$ orbitals.
Since we are interested in interpreting the XMCD measurements of Ref.~\onlinecite{hernandez2021}, we make sure that our unit cells enable the theoretically~\cite{peralta2009,chen2009,wang2014} and experimentally~\cite{moser2015} observed antiferromagnetic stripe ordering within the CuO layers.  
The unit cell is doubled in the (x,y) plane (see Fig.~\ref{fig:vacancy_struct}) when inserting an O vacancy.
To keep the computations tractable we restrict our model to a 4 CuO layers coverage.
We refer to each layer following the nomenclature of Fig.~\ref{fig:bare_interface}(a): CuO-$\{1,2,3,4\}$, SrO-$\{1,2\}$ and TiO$_2$-$\{1,2\}$.
It was shown that the experimental density of states (DOS) is best reproduced using hybrid DFT with a 8 CuO layers coverage~\cite{franchini2011}, so we chose to set the interlayer distances (up to the $4^{th}$ CuO layer) according to the values obtained by Franchini~$\etal$~\cite{franchini2011}.
We use a lateral lattice parameter $a = 3.9\si{\angstrom}$ for STO, as was obtained with hybrid DFT~\cite{wahl2008,becke1993} in excellent agreement with the experiments~\cite{cao2000}. 

We first perform ionic relaxation calculations using the Vienna \emph{ab initio} simulation package (VASP)~\cite{kresse1993, *kresse1994, kresse1996a, kresse1996b} with a $6\times6\times1$ Monkhorst-Pack grid, until the maximum force is smaller than 10$\si{meV}$ on each atom. 
The CuO layers are relaxed while the STO atoms are kept fixed except when considering an oxygen vacancy in the TiO$_2$-2 interface layer, in which case the whole structure is optimized. 
We use PAW-PBE pseudo-potentials~\cite{kresse1999} with a cut-off energy of 400$\si{eV}$ for the bare interface, which is increased up to 600$\si{eV}$ for the larger unit cells with vacancies.
Then, we extract all the results presented in this paper from all-electron full-potential PBE+U calculations using the Wien2k package~\cite{wien2k, blaha2020} with a $6\times6\times1$ Monkhorst-Pack grid done on the relaxed structure as obtained from VASP. 
Due to computational limitations we controlled the size of the planewave basis set by reducing the RKmax value when necessary (the smallest value being 5.14), and we checked that limiting the basis set size did not have a significant impact on the DOS.
The Wannier fits were obtained using the \textsf{wannier90}~\cite{mostofi2008,moser2014} package. 

\begin{figure}
	\includegraphics[width=0.9\linewidth]{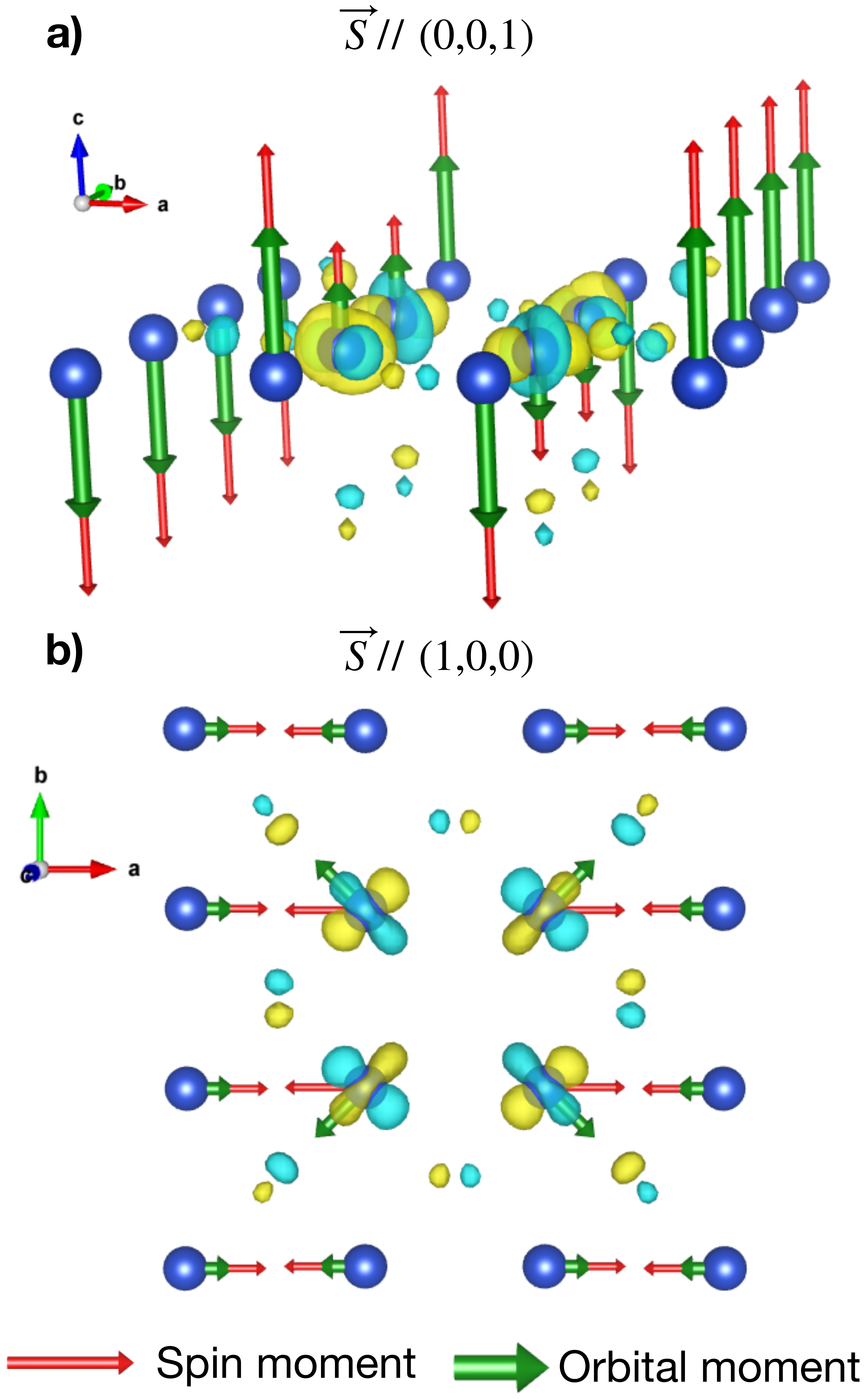}
	\caption{Schematic of the orbital moment orientation in the CuO-1 layer for two different magnetization axis: (a) (0,0,1), (b) (1,0,0). The length of the vectors illustrates the variation of the moments absolute value.}
	\label{fig:vac_VacCuO_pinnnedL}
\end{figure}

\section{Results}

\subsection{Bare Interface}

From previous studies on the t-CuO/STO junction we know that the bare (i.e. defect-free) interface should not yield a 2DEG~\cite{franchini2011,drera2019}. 
This is consistent with our results, see Fig.~\ref{fig:bare_interface}(a), in which we plot the layer-projected DOS. 
The gap $\Delta_{DFT+U}\simeq1.5 \si{eV}$ in t-CuO is layer independent and in good agreement with previous DFT+U calculations \cite{drera2019}, although smaller than the photoemission lower bound $\Delta_{ARPES}=2.4 \si{eV}$~\cite{moser2014}. 
The situation is different in the STO substrate where a clear difference can be seen between the interface and \emph{bulk} TiO$_2$ layers: the gap is reduced from $2 \si{eV}$ in the \emph{bulk} (TiO$_2$-1) to $1.5 \si{eV}$ at the interface (TiO$_2$-2).
The states reducing the gap are traced back to be of O $p_x$/$p_y$ origin. 
Fig.~\ref{fig:bare_interface}(b) shows that the $p_x$/$p_y$ orbital states are pushed towards the Fermi level at the interface, whereas the $p_z$ orbital states are stable in energy.
This is consistent with previous first principles calculations performed for TiO$_2$-terminated STO and BaTiO$_3$, and is the consequence of the emergence of pure O $p_x$/$p_y$ states not hybridized with the Ti orbitals at the surface~\cite{padilla1997,padilla1998}. 
Interestingly, adding a thin film of t-CuO above STO does not change this behavior. 
Due to the large original band gap of STO the system remains insulating since no 2DEG is observed at the bare interface, and the antiferromagnetic stripe order in t-CuO is not disturbed. 

In the following we explore the effects of inserting O vacancies in the interface layers.
We performed calculations for several possible configurations, see Fig.~\ref{fig:vacancy_struct}, some of which resulting in a scenario suitable for explaining the XMCD measurements of Ref.~\onlinecite{hernandez2021}.

\subsection{Oxygen vacancy in the interface CuO layer}

\begin{figure*}[tb]
	\includegraphics[width=\linewidth]{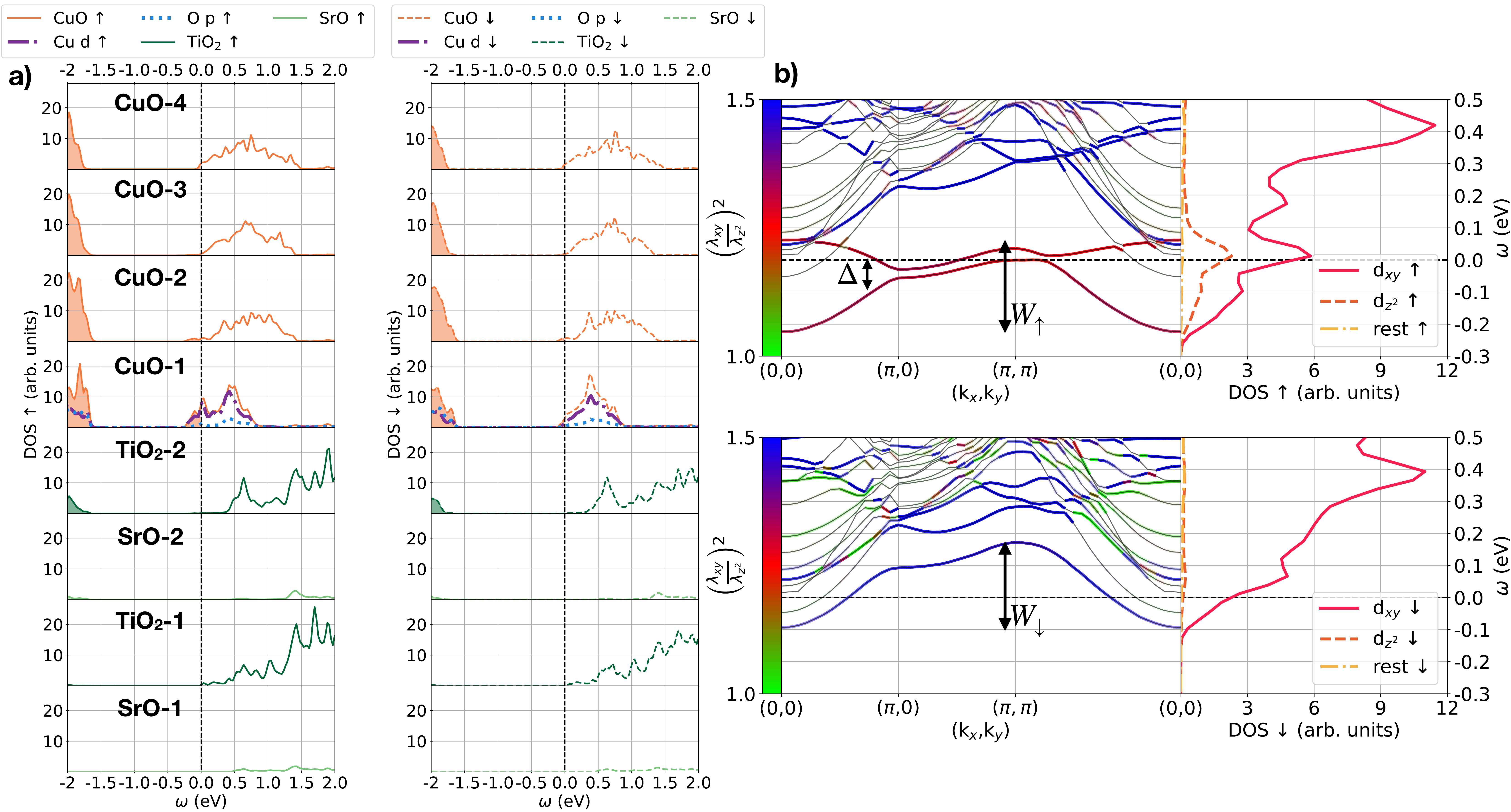}
	\caption{(a) Layer-resolved DOS for up (\emph{left}) and down (\emph{right}) spins in the case of a vacancy in the TiO$_2$-2 layer. (b) Projected band structure and DOS on the Cu $d_{xy}$ and $d_{z^2}$ orbitals of the interface CuO-1 layer. The band color depicts the ratio between the contributions from $d_{xy}$ and $d_{z^2}$, while the opacity indicates their overall contribution to the bands. The black arrows indicate the bandwidth $W_{\uparrow/\downarrow}$.}
	\label{fig:vac_TiO_DOSBANDS}
\end{figure*}

There are two geometrically different ways of creating an O vacancy in the interface CuO-1 layer (see Fig.~\ref{fig:vacancy_struct}(a)): either by removing an O having a neighboring Ti in the layer below, or an O having no such neighbor.
It is important to note that these two cases preserve the global $C_4$ symmetry both in the CuO-1 and the TiO$_2$-2 layers. 
Since the two configurations lead qualitatively to the same results, we only present here, in Fig.~\ref{fig:vac_CuO}, the ones obtained considering the vacancy without neighboring Ti atom. 
For this type of vacancy, STO behaves similarly to the bare interface configuration, see Fig.~\ref{fig:vac_CuO}(a). 
The gap is again reduced by $0.5 \si{eV}$ at the interface, but with the addition of small extra contributions around $\pm 0.4 \si{eV}$ sparked by the large in-gap states appearing in the interface CuO-1 layer. 
The latter shows a drastic change compared to the bare interface case since a new set of in-gap states of mostly Cu $d$ orbital character appears. 
These states are localized around the oxygen vacancy, as can be seen in Fig.~\ref{fig:vac_CuO}(b) where we show the projected DOS on the nearest-neighbour (NN), next-NN (NNN) and next-NNN (NNNN) Cu atoms. 
The fact that only the four NN have sizable contribution, that the oxygen contribution is weak, and that the surrounding layers' DOS display a vanishing density at $\pm0.4\si{eV}$ show the localized nature of the in-gap states both in- and out-of-plane.
It is the $d_{xy}$ orbital of the NN Cu atoms surrounding the vacancy that mostly contributes, see Fig.~\ref{fig:vac_CuO}(c). 

Therefore, despite the large value of the Hubbard interaction U, it is energetically favorable for the additional electrons to remain inside the CuO-1 layer: When a vacancy is formed, an extra electron stemming from the missing O occupies the Cu $d$ orbitals of the upper Hubbard band instead of migrating to the TiO$_2$-2 layer, where it would occupy the Ti $d$ shell. 
This is crucial when a vacancy is formed in the TiO$_2$-2 layer, as we shall describe later in this paper.

The missing O not only leads to an increase of the electronic density on the surrounding NN Cu sites, but also to the breaking of the local $C_4$ symmetry.
This is clearly seen in the spin density plot of Fig.~\ref{fig:vac_CuO}(d): the local $C_4$ symmetry on each NN Cu site hosting the extra charges is broken.
The symmetry of the in-gap states around the Fermi level and the position of the Cu atoms around the vacancy advocate the formation of bonding/anti-bonding localized states around the vacancy, thus preventing the formation of an electron gas.
It is confirmed by a fit of the in-gap bands with Wannier functions, as shown in Fig.~\ref{fig:vac_CuO}(e).
Four Wannier functions centered on the NN Cu atoms were used for the fit, i.e. two Cu for each spin flavor, which matches perfectly the two PBE+U bands. 
Consistently with the spin density, the Wannier orbitals break the local $C_4$ symmetry, and they expand towards the vacancy position in the middle of the four NN Cu. 
When deriving a tight-binding Hamiltonian from these Wannier orbitals, this change translates into a large hopping amplitude between the NN Cu sites inside the unit-cell $t_{in}\simeq0.35\si{eV}$.
Moreover, since the states are localized around the vacancy the hopping outside the unit cell is small: $t_{out}\simeq t_{in}/10$.
Hence, the splitting between the two bands $\Delta\simeq2t_{in}$ is consistent with a dimerization picture, and large compared to the small bandwidth $W$ since $t_{out}$ is one order of magnitude smaller than $t_{in}$.

Alike the previous case, the system remains insulating (although the gap is reduced at the interface CuO-1 layer) and antiferromagnetic, i.e. no paramagnetic spin moments would be observed.
Still, it is interesting to study the orbital moments obtained from the calculations including the SOC. 

Indeed, as is advocated in Ref.~\onlinecite{audehm2016}, a crystalline defect could be at the origin of orbital moment pinning. 
If the magnetization axis is set out-of plane (0,0,1), the orbital moment is parallel and proportional to the local spin, such that on the NN Cu sites the absolute value of the orbital moment is reduced, as illustrated in Fig.~\ref{fig:vac_VacCuO_pinnnedL}(a). 
On every site the magnetization is $m_{spin}\simeq\pm0.70\mu_B$ and $m_{orb}\simeq\pm0.14\mu_B$, while on the four NN sites it is weaker: $m_{spin}\simeq\pm0.45\mu_B$ and $m_{orb}\simeq\pm0.09\mu_B$.
Moreover, as the four NN sites are equally dispatched on the up and down spin stripes the total orbital moment is zero. 
Most importantly, the out-of plane part of the orbital moment is collinear with the spin. 

The situation is different when setting the magnetization axis in-plane, see Fig.~\ref{fig:vac_VacCuO_pinnnedL}(b). 
If it is along (1,0,0), or (0,1,0), the orbital moment is collinear with the spin on all sites but the four NN ones, where it is pinned along ($\pm$1,$\pm$1,0).
Moreover, its value is not proportional anymore to the spin moment, it is even inversely proportional if the magnetization axis is set along (1,1,0) for the two NN atoms aligned along (1,1,0) around the vacancy. 
Indeed, in such a case, most of the sites display a spin moment $m_{spin}\simeq\pm0.70\mu_B$ and $m_{orb}\simeq\pm0.05\mu_B$ (along the magnetization direction), unless for the two aforementioned sites which carry $m_{spin}\simeq\pm0.45\mu_B$ and $m_{orb}\simeq\pm0.07\mu_B$.
Therefore, although the total orbital moment averages out to zero because the global $C_4$ symmetry is not broken around the vacancy, we identify here a pinning mechanism which is intimately linked with the breaking of the on-site, i.e. local, $C_4$ symmetry. 
On the four NN sites, the in-plane part of the orbital moment can be non-collinear and its absolute value even inversely proportional to that of the spin moment.
This can be understood in terms of a competition between the vacancy-induced crystalline anisotropy and the SOC~\cite{durr1996}.
The in-plane anisotropy is large on the four Cu sites surrounding the vacancy, and dominates the weak SOC of Cu (which would promote the alignment of the orbital and spin moments), making it possible to obtain non-collinear magnetic moments in-plane. 

\vspace{-0.005cm}
Hence, while this scenario is not suited for explaining the paramagnetic spin moment feature, it provides precious insights on the orbital moment behavior. 
A key argument in favor of this pinning mechanism is that it would be consistent with the XMCD measurements in which only the in-plane component is pinned. 
None of these conclusions change when considering an O vacancy lying above a Ti site instead of vacuum. 

\subsection{Oxygen vacancy in the interface TiO$_2$ layer}

The physics is different if the O vacancy is created in the TiO$_2$-2 layer instead (Fig.~\ref{fig:vacancy_struct}(b)).
The layer- and spin-resolved DOS are presented in Fig.~\ref{fig:vac_TiO_DOSBANDS}(a): the CuO-1 layer at the interface is metallic and spin polarized.
Despite the global shift of the chemical potential, the system is only metallic at the interface since sizable DOS at the Fermi level only appears there.
Similarly to the previous case, those states are mostly of Cu $d$ orbital character with almost no contribution from the O $p$ orbitals.  
A striking result is that the DOS of the interface TiO$_2$-2 layer, where the O vacancy is located, seems not to be affected by the defect: the valence of the Ti atoms is unchanged and they remain non-magnetic.
So upon formation of an O vacancy in the TiO$_2$-2 layer, the electrons of the missing O atom occupy states in the neighboring CuO-1 layer.
This is important, since it shows that it is possible to generate a polarized 2DEG at the interface CuO-1 layer which is invisible at the Ti $L_{2,3}$-edge. 
Moreover, we note that the O vacancy does not intrinsically favor positive spin moments.
The sign of the resulting moment depends on the defect and the stripes relative position, so that if we shift the vacancy or if we swap the up and down spin stripes we obtain a down-spin polarized electron gas. 
Therefore, on average in the real system without external magnetic field the 2DEG should not be polarized, as expected for paramagnetic spin moments.
These considerations contradict the claim in Ref.~\onlinecite{hernandez2021} that a 2DEG at the interface should be seen in the XMCD spectrum at the Ti $L_{2,3}$-edge, and restore it as a plausible explanation for the puzzling paramagnetic moments observed at the Cu $L_{2,3}$-edge. 
\begin{figure}
	\includegraphics[width=0.9\linewidth]{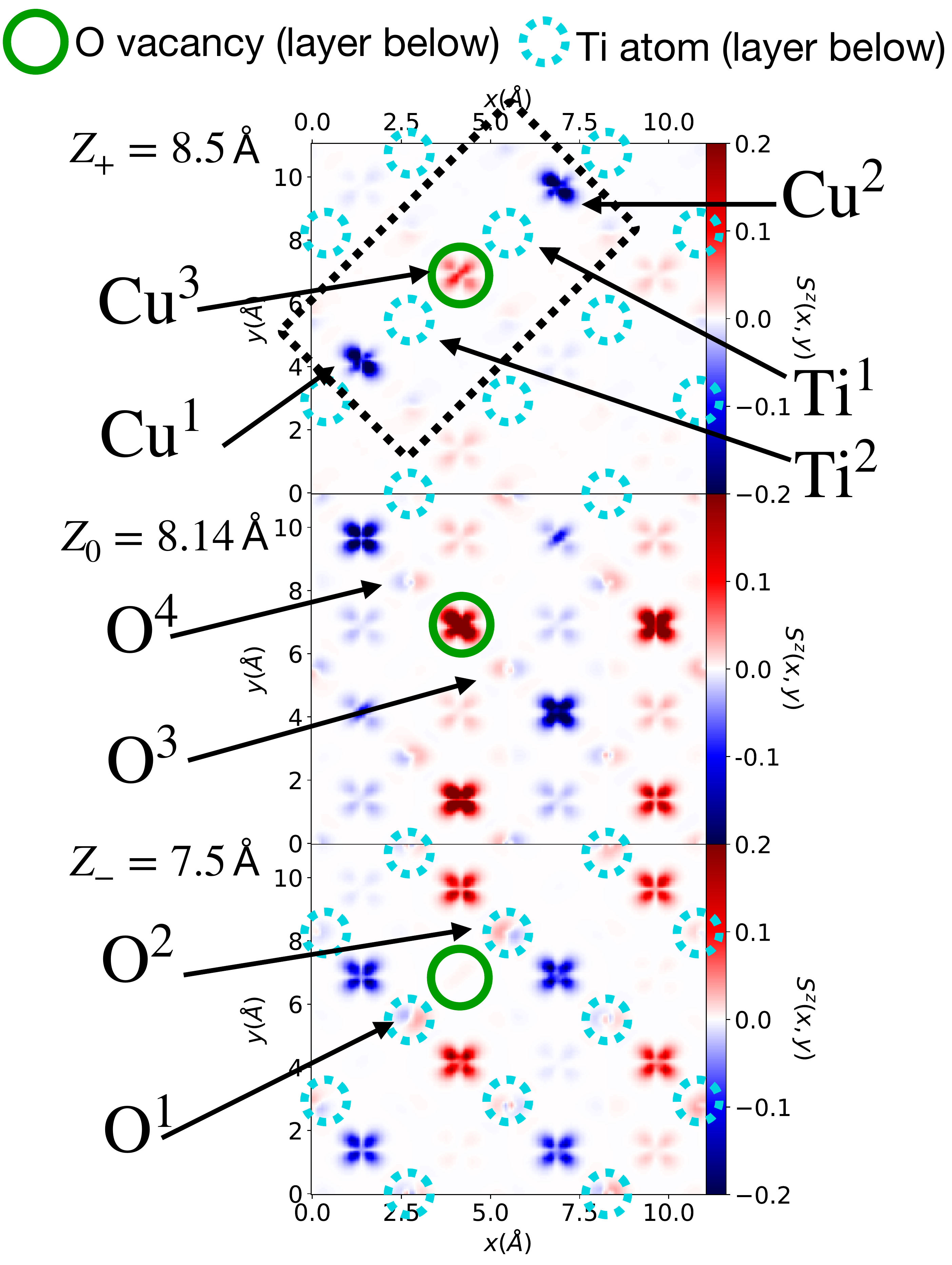}
	\caption{Spin density map of the interface CuO-1 layer below (\emph{bottom}), at (\emph{center}), and above (\emph{top}) the initial position $Z_0$ before relaxation. As a guide to the eye, the position of the O vacancy and the Ti atoms in the TiO$_2$-2 layer below is shown. The black frame highlight the three Cu atom on which is focused Fig.~\ref{fig:vac_buckling}(b).}
	\label{fig:vac_spindiff}
\end{figure}

In Fig.~\ref{fig:vac_TiO_DOSBANDS}(b) we show the spin- and orbital-projected bandstructure.
There is a striking difference between the two spin flavors: we find a single down spin $d_{xy}$ band crossing the Fermi level, whereas we observe for the up spin two split bands of a mixed Cu $d_{xy}$/$d_{z^2}$ character. 
We find the mixing and the band shape to be analogous to the SOC- and distortion-induced splitting of the Ir $d$ band in Sr$_2$IrO$_4$~\cite{lenz2019}, although it is remarkable that in our case only the up spin electrons are affected.
If the splitting of the up spin band is due to the structural distortions similarly to Sr$_2$IrO$_4$, here the band mixing however is due to the $C_4$ symmetry breaking and not the SOC. 
We note that the bands at the Fermi energy have a small bandwidth $W_{\uparrow/\downarrow}\simeq300\si{meV}$, hence being sign of a low-mobility 2DEG.

\begin{figure*}[tp]
	\includegraphics[width=0.9\linewidth]{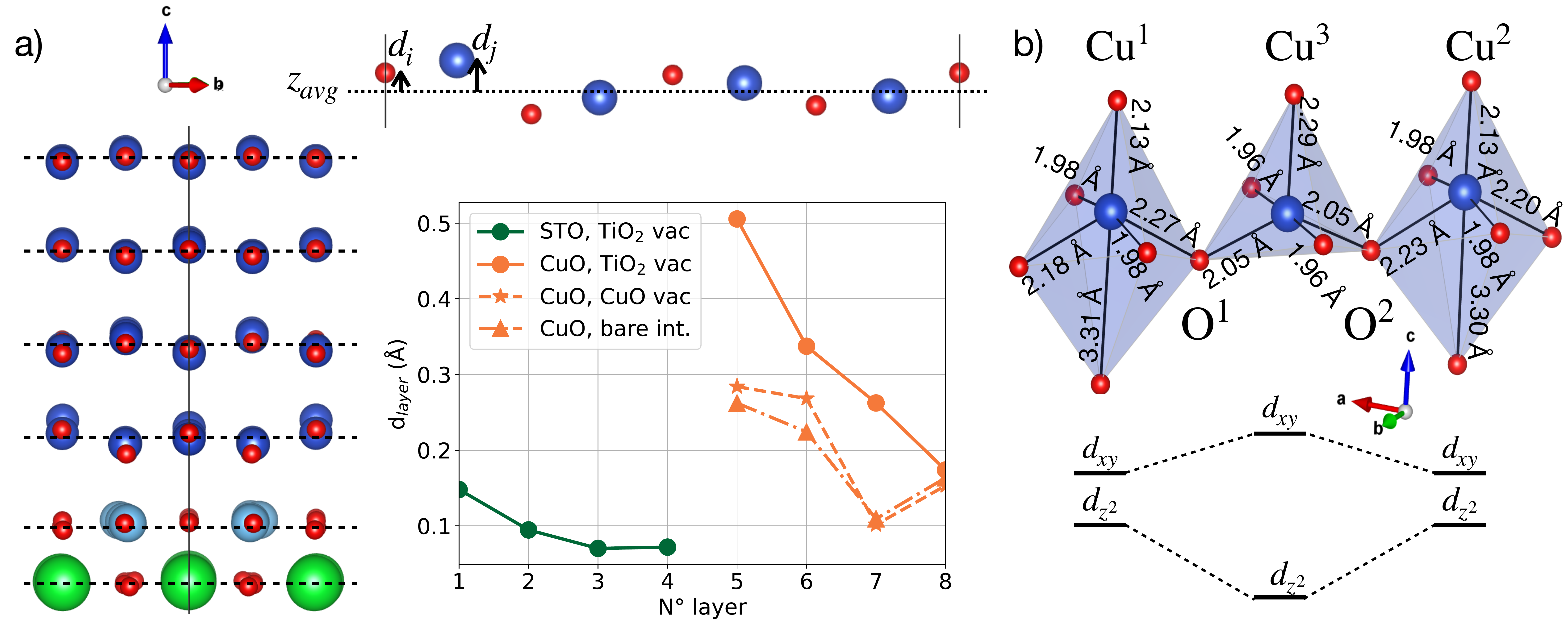}
	\caption{(a) Buckling on a portion of the crystal structure (\emph{left}), illustration of how are evaluated the displacement $d_i$ (\emph{top right}) and evaluation of the average buckling for each layer for the three first scenarios (\emph{bottom right}) following Eq.~\ref{eq:buckling}. (b) View of the three Cu atoms highlighted in Fig.~\ref{fig:vac_spindiff} with a schematic of the crystal field splitting of the $d_{xy}$ and $d_{z^2}$ orbitals.}
	\label{fig:vac_buckling}
\end{figure*} 

In Fig.~\ref{fig:vac_spindiff} we show the spin-density at the interface CuO-1 layer around its initial $Z_0$ position. 
We observe a large distortion: the spin density for the same layer is non zero over more than $1\si{\angstrom}$ along the $c$ axis.
The two Cu$^{1,2}$ atoms belonging to the down spin stripe show a breaking of the local $C_4$ symmetry due to the extra electrons coming from the TiO$_2$-2 layer below, see the \emph{top} panel.
Interestingly, these charges are not located on the Cu$^{3}$ site just above the vacancy, but rather on the Cu$^{1,2}$ sites. 
This is caused by the large buckling induced by the introduction of the vacancy, as can be seen in Fig.~\ref{fig:vac_buckling}(a). 
The average buckling along the $c$ axis per layer is evaluated as follow:
\begin{equation}
	\label{eq:buckling}
	d_{layer} = \frac{1}{N_{layer}}\sum_{i\in layer}\left|d_i\right| =\frac{1}{N_{layer}}\sum_{i\in layer}\left|z_i-z_{avg}\right|
\end{equation}
where $N_{layer}$ is the number of atom per layer, $z_i$ the $z$ position of the atom $i$, and $z_{avg}$ the average $z$ position of the layer. 
The largest buckling happens in the CuO-1 interface layer, and is of the order of 0.5$\si{\angstrom}$.
It is interesting to note that the distortions are larger than for the bare interface and the case of a vacancy in the CuO-1 layer. 
Moreover, the overall distortion follows the sublattice decoupling observed experimentally~\cite{moser2014,moser2015,hernandez2021,zhong2021} and recently explained theoretically~\cite{bramberger2022}.
Note that the upturn in the atomic displacement in STO as going towards the first layer is due to the vacuum below the last substrate layer since the whole cell is optimized. 
It remains significantly smaller than the buckling in the CuO-1 layer. 

This large buckling of the CuO-1 layer is driven by the displacement of the O$^{1,2}$ atoms located above the two Ti$^{1,2}$ sites surrounding the vacancy, which can be understood within a simple electrostatic picture. 
Upon formation of an O vacancy, the valence of the Ti$^{1,2}$ atoms could change to Ti$^{3+}$ and the TiO$_2$-2 layer would remain neutral. 
However, since even with strong electron-electron interaction U the system favors the migration of the additional electrons into the CuO-1 layer, the Ti atoms remain in a Ti$^{4+}$ valence.
This leads to an overall local polarization of the two interface layers around the vacancy: TiO$_2$-2$^{2+}$ and CuO-1$^{2-}$.
To minimize the electrostatic energy, the O$^{1,2}$ atoms (charged negatively) are pulled down towards the Ti$^{1,2}$ atoms, hence triggering a large buckling in the CuO-1 layer.

Such displacement induces large changes in the Cu-O bonds, especially around the Cu$^{3}$ site located above the vacancy, see Fig.~\ref{fig:vac_buckling}(b). 
Since Cu$^{3}$ has no O neighbor below and its in-plane Cu-O distances are smaller than for the two surrounding Cu$^{1,2}$, the splitting between $d_{xy}$ and $d_{z^2}$ orbitals is expected to be enhanced. 
The $d_{xy}$ energy in the two neighboring Cu$^{1,2}$ sites is lower because their Cu-O distance is larger, while the $d_{z^2}$ is higher in energy because of the O below. 
Therefore, in a simple one-particle band picture the extra charges from the TiO$_2$-2 layer would occupy the Cu$^{1,2}$ sites belonging to the down-spin stripe since they are more favorable in energy.
The Cu$^{1,2}$ atoms are two of the four NNN of Cu$^{3}$, and it is not trivial why those two sites would prevail the two other NNN and even the four NN Cu sites.
In fact, all the other sites remain closer to their O neighbors, i.e. their Cu-O bonds are shorter than the ones of Cu$^{1,2}$, which implies that the crystal-field contribution favors the two Cu$^{1,2}$ sites. 
Since all the $d$ orbitals but $d_{xy}$ are filled, and the $d_{xy}$ is already occupied with a down spin electron, the extra electrons have an up spin such that the electron gas is polarized. 
The breaking of the local $C_4$ symmetry at the interface induces the mixing of the $d_{xy}$ and $d_{z^2}$ orbitals, which are furthermore closer in energy on these two sites. 
Finally, the splitting $\Delta$ of the bands in the up spin DOS is due to a dimerization: the neighboring Cu$^{1,2}$ sites contribute equally to the band above and below the Fermi energy, in contrast to the Cu$^3$ site above the vacancy which solely contributes to the single down spin band crossing the Fermi energy.
$\Delta$ is however significantly smaller that in previous case: a Wannierization of the up spin bands using only two Wannier functions centered on the Cu$^{1,2}$ sites gives an intra-unit-cell hopping $t_{in}\simeq0.035\si{eV}$, and a inter-unit-cell hopping of the same order of magnitude $t_{out}\simeq t_{in}$.

An important difference with the case of a vacancy in the CuO-1 layer is that in the latter case the $C_4$ symmetry around the vacancy is conserved (although locally it is broken on the NN Cu site). 
Here, however, the global $C_4$ is broken as can be seen in the \emph{bottom} panel of Fig.~\ref{fig:vac_spindiff}: among the four O$^{1,2,3,4}$ atoms located in the CuO-1 layer around the vacancy position, only two of them (O$^{1,2}$) are located above Ti sites. 
Those two atoms are pulled down during the relaxation process, weakening the Cu-O-Cu bond between the three Cu$^{1,2,3}$ sites as is discussed above.
Hence, in contrast with the previous case, the four NNN Cu sites around the vacancy position (note that the vacancy is in the layer below) are not equivalent anymore, and the extra electrons are hosted by only two of them (Cu$^{1,2}$), leading to a breaking of the global unit cell $C_4$ symmetry.

\begin{figure}[bp]
	\includegraphics[width=0.8\linewidth]{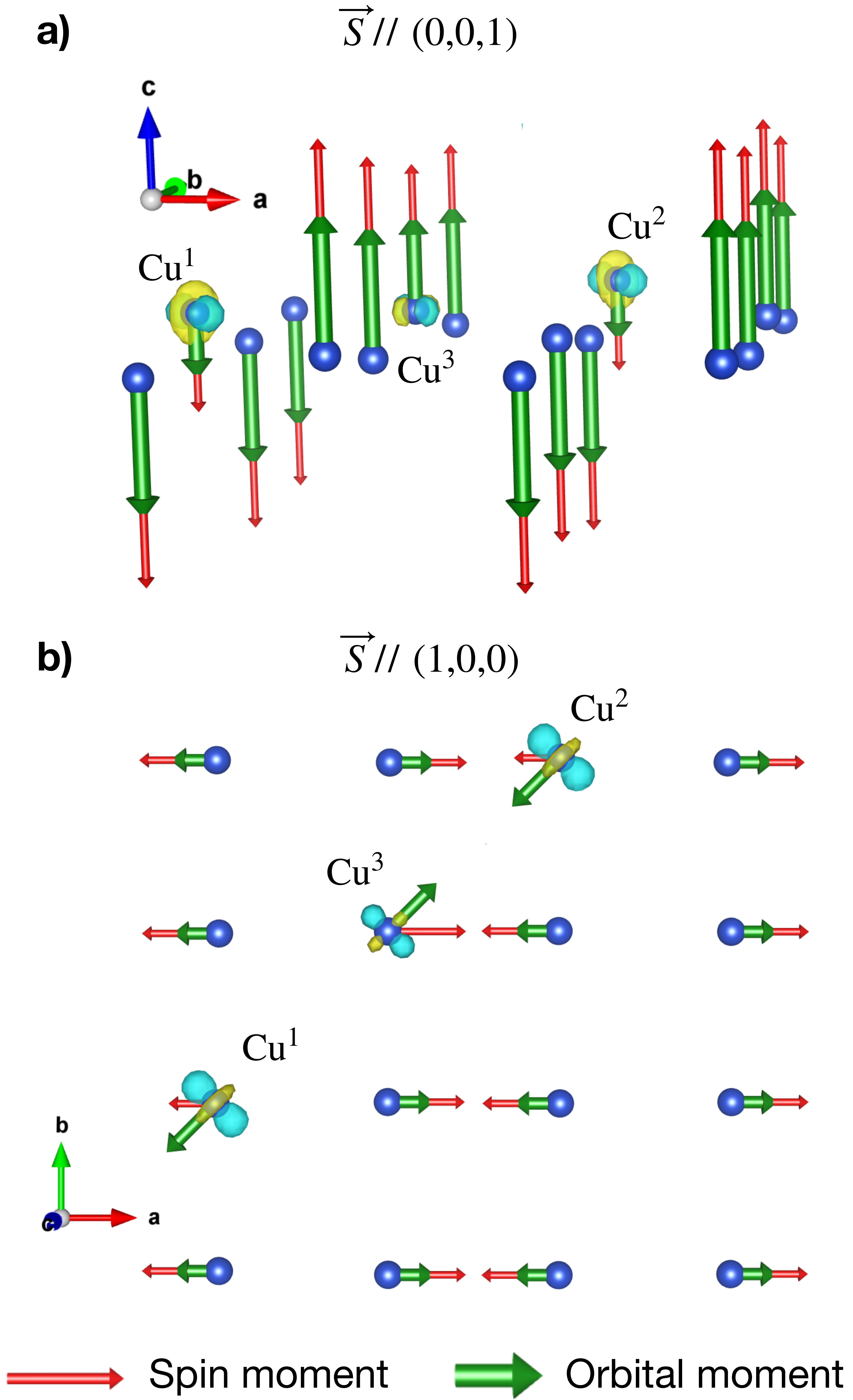}
	\caption{Schematic of the orbital moment orientation in the CuO-1 layer for two different magnetization axis: (a) (0,0,1), (b) (1,0,0). The length of the vectors illustrates the variation of the moments absolute value. For the sake of clarity the isovalue of the Wannier orbital was set to relatively high value.}
	\label{fig:vac_VacTiO_pinnnedL}
\end{figure}

If the magnetization points out of plane (Fig.~\ref{fig:vac_VacTiO_pinnnedL}(a)), so is the orbital moment and it is \emph{blind} to the in-plane symmetry breaking, such that the system is equivalent to the case where a vacancy is inserted in the CuO-1 layer (Fig.~\ref{fig:vac_VacCuO_pinnnedL}(a)). 
Indeed, the orbital moment remains proportional and collinear to the spin one, and is therefore reduced on the two Cu$^{1,2}$ sites: on most of the sites $m_{spin}\simeq\pm0.70\mu_B$ and $m_{orb}\simeq\pm0.14\mu_B$, while on the two NNN ones $m_{spin}\simeq-0.36\mu_B$ and $m_{orb}\simeq-0.08\mu_B$.
This leads to a non-zero out-of-plane orbital moment in the unit-cell, as illustrated in Fig.~\ref{fig:vac_VacTiO_pinnnedL}(a), where are also plotted the Wannier orbitals obtained by a fit of the two split spin up bands, as well as the single down-spin band.
However, since the out-of-plane component of the orbital moment remains collinear with its spin counterpart, on average in the real material without external field both components should average to zero, i.e. there is no out-of-plane orbital moment pinning.

If the magnetization is in-plane instead, see Fig.~\ref{fig:vac_VacTiO_pinnnedL}(b), we observe again a pinning of the in-plane component of the orbital moment on each of the three aforementioned sites.
In our configuration, the two Cu$^{1,2}$ sites are aligned along (1,1,0), such that the magnetization axis along (1,1,0) and (1,-1,0) are no longer equivalent (contrary to the case of a vacancy in the CuO layer, see Fig.~\ref{fig:vac_VacCuO_pinnnedL}).
We observe that, if the magnetization axis is chosen to be along (1,1,0), the orbital moment is increased in absolute value on the Cu$^{1,2}$ ($m_{orb}\simeq-0.09\mu_B$ instead of  $m_{orb}\simeq\pm0.05\mu_B$ on the other sites) although they have a reduced spin-moment with respect to the other sites ($m_{spin}\simeq-0.36\mu_B$ instead of  $m_{spin}\simeq\pm0.70\mu_B$). 
Again, if the magnetization axis is along (1,0,0) or (0,1,0), the orbital moment on all Cu atoms is collinear with the spin moment, unless for the Cu$^{1,2}$ atoms for which the orbital moment is pinned along (-1,-1,0), as well as the Cu$^{3}$ site above the vacancy for which the orbital moment is pinned along (1,1,0).
The sum of the three pinned moments is non-zero in our configuration.

Therefore, we have shown that a 2DEG appears in the CuO-1 interface layer when inserting a vacancy in the TiO$_2$-2 layer, which is invisible to XMCD at the Ti $L_{2,3}$-edge.
Around the vacancy, the orbital moment is locally pinned along (1,1,0) direction and can be non-collinear with the spin moments, with an overall non-zero value. 
The same effect could be seen in a (1,-1,0) configuration, in which case the orbital moment would be pinned along (1,-1,0) direction. 
\begin{figure}
	\includegraphics[width=\linewidth]{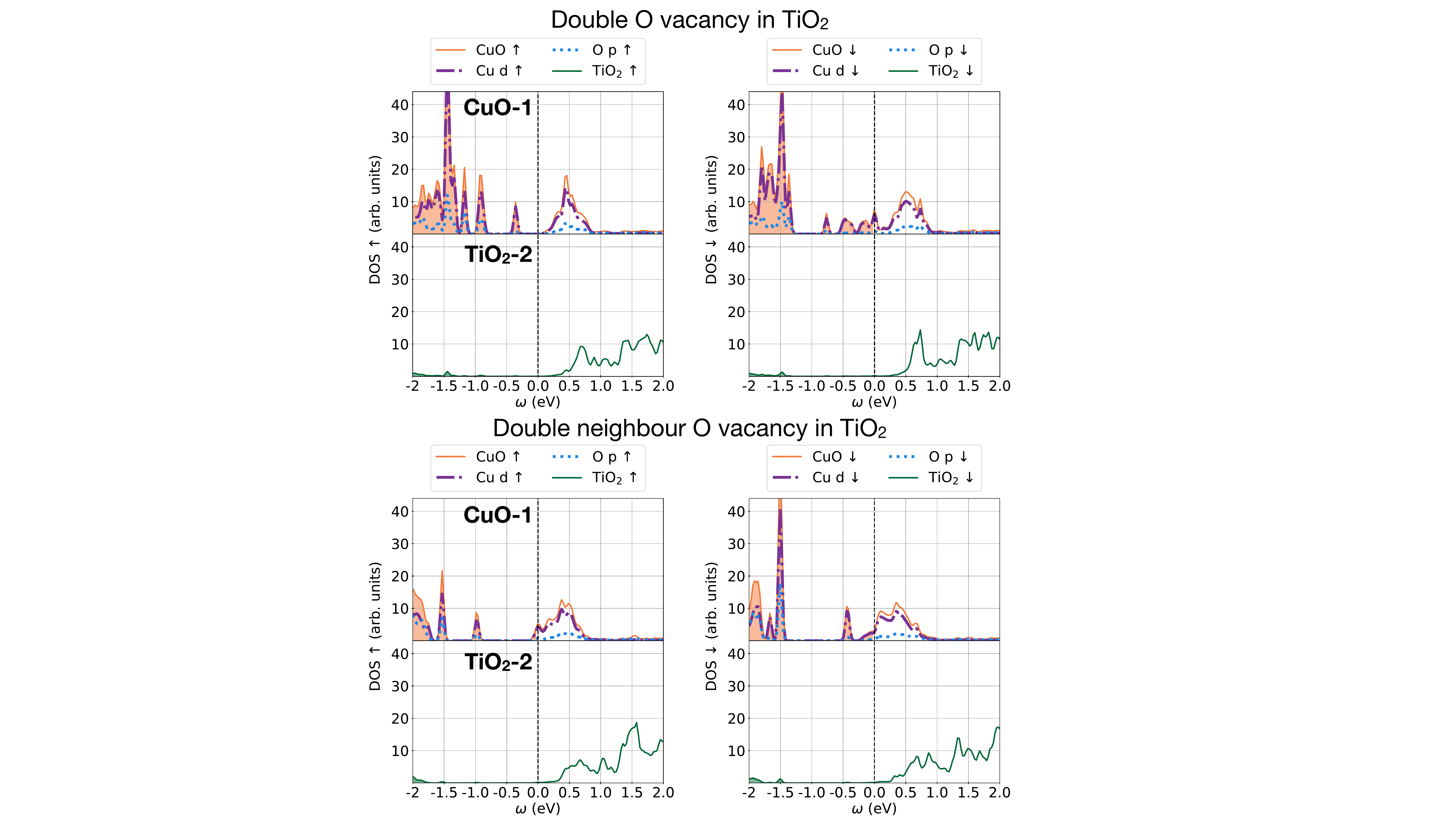}
	\caption{Layer-resolved DOS for the two interface layers in case of higher concentration of vacancies.}
	\label{fig:2vac_TiO}
\end{figure}
The striking difference here is that one vacancy gives rise to a non-zero pinned orbital moment, and it is possible to consider a distribution of vacancies resulting in an average zero magnetization but non-zero orbital moment.
Such a situation is forbidden in the case of a vacancy in the CuO-1 layer since the orbital moment is locally averaged to zero because of the global $C_4$ symmetry conservation.
This is a key feature of this scenario that should apply on similar types of monoxide/dioxide junctions: inserting a vacancy in the dioxide interface layer induces a $C_4$ symmetry breaking in the monoxide layer above which could be at the origin of a non-zero pinned orbital moment. 
Finally, as mentioned in the previous section, it is only the in-plane part of the orbital moment which is pinned, whereas its out-of-plane part is collinear and proportional to the spin moment, in excellent agreement with the XMCD measurements finding no out-of-plane pinning. 

\subsection{Two oxygen vacancies in the interface TiO$_2$ layer}

An important question is whether this scenario still holds at a higher concentration of vacancies in the TiO$_2$-2 layer. 
For instance in SrVO$_3$, two neighboring oxygen vacancies, instead of a single one are needed to reproduce the additional spectral weight seen in photoemission~\cite{backes2016}. 
Fig.~\ref{fig:2vac_TiO} shows the layer- and spin-resolved DOS obtained for a larger concentration of oxygen vacancies with two different configurations (see Fig.~\ref{fig:vacancy_struct}(c)): either two vacancies still far from each other, or forming a pair around a single Ti atom. 
Both configurations induce an electron gas in the interface CuO-1 layer while the TiO$_2$-2 layer is still insulating. 
Since we doubled the number of vacancies, there is an additional charge transfer to the interface CuO-1 layer which translates into a larger DOS below the Fermi level. 
Similarly to the previous section, the additional states at the Fermi level are almost purely of Cu $d$ character.

However, when the vacancies form a pair around the same Ti atom, the latter is not able to totally transfer the excess charges and becomes slightly spin-polarized ($m_{spin}\simeq-0.42\mu_B$), contrary to the case in which the vacancies are still apart from each other.
Such a spin moment on the Ti atoms is not consistent with the XMCD measurements, which allows us to discard the scenario of oxygen vacancies pairing around the same Ti atom. 
In contrast, when the two vacancies do not combine around the same Ti atom, the 2DEG remains invisible at the Ti $L_{2,3}$-edge, and we observe an in-plane pinning of the orbital moment in the CuO-1 layer.

\section{Conclusion}

To summarize, we present a scenario which explains the puzzling XMCD measurements of Ref.~\onlinecite{hernandez2021}. 
By performing DFT+U calculations, we show that the paramagnetic spin and pinned orbital moments observed in experiment would emerge at the t-CuO/STO interface upon the formation of oxygen vacancies. 
When located in the TiO$_2$-2 interface layer, a vacancy can induce a 2DEG in the CuO-1 layer. 
Most importantly, we observe no valence change nor non-zero magnetic moments at the interface Ti sites, which is consistent with the XMCD measurements~\cite{hernandez2021}.
The 2DEG behaves paramagnetically, and would therefore be at the origin of the paramagnetic spin moments observed experimentally.  
We identify a $C_4$ symmetry breaking in the CuO-1 layer above the vacancy as being a possible mechanism for the emergence of an in-plane pinned orbital moment, in excellent agreement with the nature of the pinned moment observed in XMCD.
We moreover show that if the oxygen vacancies form pairs around a single Ti atom, the latter becomes spin polarized and is therefore visible in XMCD.

In order to test the validity of our scenario, we propose complementary experiments that could be performed on the t-CuO/STO junction:
1) A local probe like STM would be a good candidate in order to study specifically the interface layers since it is able to spatially resolve the charge density distribution.
If applied to vertical cuts of the t-CuO/STO heterostructure as was done in Ref.~\onlinecite{samal2014}, STM would be sensitive to variations in the electronic density at the interface.
2) Since t-CuO was successfully synthesized up to 10 unit cells~\cite{zhong2021}, it may be possible to use the small probing depth of XMCD in total electron yield to suppress, or enhance, the interface contribution to the spectrum, as was done in Ref.~\onlinecite{lee2013} for LaAlO$_3$/STO.
3) Similarly, hard x-ray photoelectron spectroscopy at the Cu L$_{2,3}$-edge could be used to probe the presence of the 2DEG at the interface by studying the angle dependence of the photoemission spectrum, as was done for LaAlO$_3$/STO~\cite{sing2009}.
4) Instead, the 2DEG could also be suppressed at the interface by using a metallic Nb-doped STO substrate, which was already used as a substrate for t-CuO~\cite{moser2015}, or by changing the termination of STO as was done for instance for AlO$_2$/STO interfaces~\cite{mannhart2008}.

Since the orientation of the magnetization axis plays an important role in the emergence of the pinned orbital moment, its experimental determination will be an important step to advance the understanding of magnetic properties of t-CuO. 

Finally, we would like to stress that our results may have implications to other systems.
Not only the O vacancies play an important role in many oxides~\cite{gunkel2020}, but also the pinning of the orbital moment has been observed in other oxides~\cite{buchner2018} and exchange-bias systems~\cite{audehm2016}. 
Our demonstration that the orbital moment pinning may be sparked by structural distortions breaking the local and global $C_4$ rotation symmetries could therefore apply to other systems, too.
This is especially true for other monoxide/dioxide junctions with similar metal-oxygen bond length, and weak SOC.
At the interface, the oxygen atoms of the monoxide layer would separate into two groups, those lying above an empty site and those above a metallic ion (as is the case for t-CuO/STO).
As we have demonstrated, such separation can spark a symmetry breaking when a defect, like a vacancy, is inserted, and thus can lead to the pinning of the orbital moment. 

\section*{Acknowledgments}
We thank Andr\'es Felipe Santander-Syro, Simon Moser, Steffen Backes, Max Bramberger, Martin Grundner and Sebastian Paeckel for fruitful discussions.
B.B.-L. acknowledges funding through the Institut Polytechnique de Paris.
We acknowledge supercomputing time at IDRIS-GENCI Orsay (Project No. A0110901393) and we thank the CPHT computer support team.

\section*{Data Availability}
Source data for figures is available from the corresponding authors upon reasonable request.

\bibliographystyle{apsrev4-2}

\bibliography{references.bib}

\begin{thebibliography}{54}%
\makeatletter
\providecommand \@ifxundefined [1]{%
 \@ifx{#1\undefined}
}%
\providecommand \@ifnum [1]{%
 \ifnum #1\expandafter \@firstoftwo
 \else \expandafter \@secondoftwo
 \fi
}%
\providecommand \@ifx [1]{%
 \ifx #1\expandafter \@firstoftwo
 \else \expandafter \@secondoftwo
 \fi
}%
\providecommand \natexlab [1]{#1}%
\providecommand \enquote  [1]{``#1''}%
\providecommand \bibnamefont  [1]{#1}%
\providecommand \bibfnamefont [1]{#1}%
\providecommand \citenamefont [1]{#1}%
\providecommand \href@noop [0]{\@secondoftwo}%
\providecommand \href [0]{\begingroup \@sanitize@url \@href}%
\providecommand \@href[1]{\@@startlink{#1}\@@href}%
\providecommand \@@href[1]{\endgroup#1\@@endlink}%
\providecommand \@sanitize@url [0]{\catcode `\\12\catcode `\$12\catcode
  `\&12\catcode `\#12\catcode `\^12\catcode `\_12\catcode `\%12\relax}%
\providecommand \@@startlink[1]{}%
\providecommand \@@endlink[0]{}%
\providecommand \url  [0]{\begingroup\@sanitize@url \@url }%
\providecommand \@url [1]{\endgroup\@href {#1}{\urlprefix }}%
\providecommand \urlprefix  [0]{URL }%
\providecommand \Eprint [0]{\href }%
\providecommand \doibase [0]{https://doi.org/}%
\providecommand \selectlanguage [0]{\@gobble}%
\providecommand \bibinfo  [0]{\@secondoftwo}%
\providecommand \bibfield  [0]{\@secondoftwo}%
\providecommand \translation [1]{[#1]}%
\providecommand \BibitemOpen [0]{}%
\providecommand \bibitemStop [0]{}%
\providecommand \bibitemNoStop [0]{.\EOS\space}%
\providecommand \EOS [0]{\spacefactor3000\relax}%
\providecommand \BibitemShut  [1]{\csname bibitem#1\endcsname}%
\let\auto@bib@innerbib\@empty
\bibitem [{\citenamefont {Hwang}\ \emph {et~al.}(2012)\citenamefont {Hwang},
  \citenamefont {Iwasa}, \citenamefont {Kawasaki}, \citenamefont {Keimer},
  \citenamefont {Nagaosa},\ and\ \citenamefont {Tokura}}]{hwang2012}%
  \BibitemOpen
  \bibfield  {author} {\bibinfo {author} {\bibfnamefont {H.~Y.}\ \bibnamefont
  {Hwang}}, \bibinfo {author} {\bibfnamefont {Y.}~\bibnamefont {Iwasa}},
  \bibinfo {author} {\bibfnamefont {M.}~\bibnamefont {Kawasaki}}, \bibinfo
  {author} {\bibfnamefont {B.}~\bibnamefont {Keimer}}, \bibinfo {author}
  {\bibfnamefont {N.}~\bibnamefont {Nagaosa}},\ and\ \bibinfo {author}
  {\bibfnamefont {Y.}~\bibnamefont {Tokura}},\ }\href
  {https://doi.org/10.1038/nmat3223} {\bibfield  {journal} {\bibinfo  {journal}
  {Nature Mater}\ }\textbf {\bibinfo {volume} {11}},\ \bibinfo {pages} {103}
  (\bibinfo {year} {2012})}\BibitemShut {NoStop}%
\bibitem [{\citenamefont {Kumah}\ \emph {et~al.}(2020)\citenamefont {Kumah},
  \citenamefont {Ngai},\ and\ \citenamefont {Kornblum}}]{kumah2020}%
  \BibitemOpen
  \bibfield  {author} {\bibinfo {author} {\bibfnamefont {D.~P.}\ \bibnamefont
  {Kumah}}, \bibinfo {author} {\bibfnamefont {J.~H.}\ \bibnamefont {Ngai}},\
  and\ \bibinfo {author} {\bibfnamefont {L.}~\bibnamefont {Kornblum}},\ }\href
  {https://doi.org/10.1002/adfm.201901597} {\bibfield  {journal} {\bibinfo
  {journal} {Advanced Functional Materials}\ }\textbf {\bibinfo {volume}
  {30}},\ \bibinfo {pages} {1901597} (\bibinfo {year} {2020})}\BibitemShut
  {NoStop}%
\bibitem [{\citenamefont {Koster}\ \emph {et~al.}(2015)\citenamefont {Koster},
  \citenamefont {Huijben},\ and\ \citenamefont {Rijnders}}]{koster2015}%
  \BibitemOpen
  \bibfield  {author} {\bibinfo {author} {\bibfnamefont {G.}~\bibnamefont
  {Koster}}, \bibinfo {author} {\bibfnamefont {M.}~\bibnamefont {Huijben}},\
  and\ \bibinfo {author} {\bibfnamefont {G.}~\bibnamefont {Rijnders}},\
  }\href@noop {} {\emph {\bibinfo {title} {Epitaxial {{Growth}} of {{Complex
  Metal Oxides}}}}}\ (\bibinfo  {publisher} {{Elsevier}},\ \bibinfo {year}
  {2015})\BibitemShut {NoStop}%
\bibitem [{\citenamefont {Siemons}\ \emph {et~al.}(2009)\citenamefont
  {Siemons}, \citenamefont {Koster}, \citenamefont {Blank}, \citenamefont
  {Hammond}, \citenamefont {Geballe},\ and\ \citenamefont
  {Beasley}}]{siemons2009}%
  \BibitemOpen
  \bibfield  {author} {\bibinfo {author} {\bibfnamefont {W.}~\bibnamefont
  {Siemons}}, \bibinfo {author} {\bibfnamefont {G.}~\bibnamefont {Koster}},
  \bibinfo {author} {\bibfnamefont {D.~H.~A.}\ \bibnamefont {Blank}}, \bibinfo
  {author} {\bibfnamefont {R.~H.}\ \bibnamefont {Hammond}}, \bibinfo {author}
  {\bibfnamefont {T.~H.}\ \bibnamefont {Geballe}},\ and\ \bibinfo {author}
  {\bibfnamefont {M.~R.}\ \bibnamefont {Beasley}},\ }\href
  {https://doi.org/10.1103/PhysRevB.79.195122} {\bibfield  {journal} {\bibinfo
  {journal} {Phys. Rev. B}\ }\textbf {\bibinfo {volume} {79}},\ \bibinfo
  {pages} {195122} (\bibinfo {year} {2009})}\BibitemShut {NoStop}%
\bibitem [{\citenamefont {Zhong}\ \emph {et~al.}(2021)\citenamefont {Zhong},
  \citenamefont {Dou}, \citenamefont {Wang}, \citenamefont {Lv}, \citenamefont
  {Han}, \citenamefont {Yan}, \citenamefont {Song}, \citenamefont {Ma},\ and\
  \citenamefont {Xue}}]{zhong2021}%
  \BibitemOpen
  \bibfield  {author} {\bibinfo {author} {\bibfnamefont {Y.}~\bibnamefont
  {Zhong}}, \bibinfo {author} {\bibfnamefont {Z.}~\bibnamefont {Dou}}, \bibinfo
  {author} {\bibfnamefont {R.-F.}\ \bibnamefont {Wang}}, \bibinfo {author}
  {\bibfnamefont {Y.-F.}\ \bibnamefont {Lv}}, \bibinfo {author} {\bibfnamefont
  {S.}~\bibnamefont {Han}}, \bibinfo {author} {\bibfnamefont {H.}~\bibnamefont
  {Yan}}, \bibinfo {author} {\bibfnamefont {C.-L.}\ \bibnamefont {Song}},
  \bibinfo {author} {\bibfnamefont {X.-C.}\ \bibnamefont {Ma}},\ and\ \bibinfo
  {author} {\bibfnamefont {Q.-K.}\ \bibnamefont {Xue}},\ }\href
  {https://doi.org/10.1063/5.0069356} {\bibfield  {journal} {\bibinfo
  {journal} {Appl. Phys. Lett.}\ }\textbf {\bibinfo {volume} {119}},\ \bibinfo
  {pages} {172602} (\bibinfo {year} {2021})}\BibitemShut {NoStop}%
\bibitem [{\citenamefont {{\AA}sbrink}\ and\ \citenamefont
  {Norrby}(1970)}]{asbrink1970}%
  \BibitemOpen
  \bibfield  {author} {\bibinfo {author} {\bibfnamefont {S.}~\bibnamefont
  {{\AA}sbrink}}\ and\ \bibinfo {author} {\bibfnamefont {L.-J.}\ \bibnamefont
  {Norrby}},\ }\href {https://doi.org/10.1107/S0567740870001838} {\bibfield
  {journal} {\bibinfo  {journal} {Acta Cryst B}\ }\textbf {\bibinfo {volume}
  {26}},\ \bibinfo {pages} {8} (\bibinfo {year} {1970})}\BibitemShut {NoStop}%
\bibitem [{\citenamefont {Samal}\ \emph {et~al.}(2014)\citenamefont {Samal},
  \citenamefont {Tan}, \citenamefont {Takamura}, \citenamefont {Siemons},
  \citenamefont {Verbeeck}, \citenamefont {Van~Tendeloo}, \citenamefont
  {Arenholz}, \citenamefont {Jenkins}, \citenamefont {Rijnders},\ and\
  \citenamefont {Koster}}]{samal2014}%
  \BibitemOpen
  \bibfield  {author} {\bibinfo {author} {\bibfnamefont {D.}~\bibnamefont
  {Samal}}, \bibinfo {author} {\bibfnamefont {H.}~\bibnamefont {Tan}}, \bibinfo
  {author} {\bibfnamefont {Y.}~\bibnamefont {Takamura}}, \bibinfo {author}
  {\bibfnamefont {W.}~\bibnamefont {Siemons}}, \bibinfo {author} {\bibfnamefont
  {J.}~\bibnamefont {Verbeeck}}, \bibinfo {author} {\bibfnamefont
  {G.}~\bibnamefont {Van~Tendeloo}}, \bibinfo {author} {\bibfnamefont
  {E.}~\bibnamefont {Arenholz}}, \bibinfo {author} {\bibfnamefont {C.~A.}\
  \bibnamefont {Jenkins}}, \bibinfo {author} {\bibfnamefont {G.}~\bibnamefont
  {Rijnders}},\ and\ \bibinfo {author} {\bibfnamefont {G.}~\bibnamefont
  {Koster}},\ }\href {https://doi.org/10.1209/0295-5075/105/17003} {\bibfield
  {journal} {\bibinfo  {journal} {EPL}\ }\textbf {\bibinfo {volume} {105}},\
  \bibinfo {pages} {17003} (\bibinfo {year} {2014})}\BibitemShut {NoStop}%
\bibitem [{\citenamefont {Anderson}(1987)}]{anderson1987}%
  \BibitemOpen
  \bibfield  {author} {\bibinfo {author} {\bibfnamefont {P.~W.}\ \bibnamefont
  {Anderson}},\ }\href {https://doi.org/10.1126/science.235.4793.1196}
  {\bibfield  {journal} {\bibinfo  {journal} {Science}\ }\textbf {\bibinfo
  {volume} {235}},\ \bibinfo {pages} {1196} (\bibinfo {year}
  {1987})}\BibitemShut {NoStop}%
\bibitem [{\citenamefont {Zaanen}\ \emph {et~al.}(1985)\citenamefont {Zaanen},
  \citenamefont {Sawatzky},\ and\ \citenamefont {Allen}}]{zaanen1985}%
  \BibitemOpen
  \bibfield  {author} {\bibinfo {author} {\bibfnamefont {J.}~\bibnamefont
  {Zaanen}}, \bibinfo {author} {\bibfnamefont {G.~A.}\ \bibnamefont
  {Sawatzky}},\ and\ \bibinfo {author} {\bibfnamefont {J.~W.}\ \bibnamefont
  {Allen}},\ }\href {https://doi.org/10.1103/PhysRevLett.55.418} {\bibfield
  {journal} {\bibinfo  {journal} {Phys. Rev. Lett.}\ }\textbf {\bibinfo
  {volume} {55}},\ \bibinfo {pages} {418} (\bibinfo {year} {1985})}\BibitemShut
  {NoStop}%
\bibitem [{\citenamefont {Emery}(1987)}]{emery1987}%
  \BibitemOpen
  \bibfield  {author} {\bibinfo {author} {\bibfnamefont {V.~J.}\ \bibnamefont
  {Emery}},\ }\href {https://doi.org/10.1103/PhysRevLett.58.2794} {\bibfield
  {journal} {\bibinfo  {journal} {Phys. Rev. Lett.}\ }\textbf {\bibinfo
  {volume} {58}},\ \bibinfo {pages} {2794} (\bibinfo {year}
  {1987})}\BibitemShut {NoStop}%
\bibitem [{\citenamefont {Zhang}\ and\ \citenamefont {Rice}(1988)}]{zhang1988}%
  \BibitemOpen
  \bibfield  {author} {\bibinfo {author} {\bibfnamefont {F.~C.}\ \bibnamefont
  {Zhang}}\ and\ \bibinfo {author} {\bibfnamefont {T.~M.}\ \bibnamefont
  {Rice}},\ }\href {https://doi.org/10.1103/PhysRevB.37.3759} {\bibfield
  {journal} {\bibinfo  {journal} {Phys. Rev. B}\ }\textbf {\bibinfo {volume}
  {37}},\ \bibinfo {pages} {3759} (\bibinfo {year} {1988})}\BibitemShut
  {NoStop}%
\bibitem [{\citenamefont {Andersen}\ \emph {et~al.}(1995)\citenamefont
  {Andersen}, \citenamefont {Liechtenstein}, \citenamefont {Jepsen},\ and\
  \citenamefont {Paulsen}}]{andersen1995}%
  \BibitemOpen
  \bibfield  {author} {\bibinfo {author} {\bibfnamefont {O.~K.}\ \bibnamefont
  {Andersen}}, \bibinfo {author} {\bibfnamefont {A.~I.}\ \bibnamefont
  {Liechtenstein}}, \bibinfo {author} {\bibfnamefont {O.}~\bibnamefont
  {Jepsen}},\ and\ \bibinfo {author} {\bibfnamefont {F.}~\bibnamefont
  {Paulsen}},\ }\href {https://doi.org/10.1016/0022-3697(95)00269-3} {\bibfield
   {journal} {\bibinfo  {journal} {Journal of Physics and Chemistry of Solids}\
  }\bibinfo {series} {Proceedings of the {{Conference}} on {{Spectroscopies}}
  in {{Novel Superconductors}}},\ \textbf {\bibinfo {volume} {56}},\ \bibinfo
  {pages} {1573} (\bibinfo {year} {1995})}\BibitemShut {NoStop}%
\bibitem [{\citenamefont {Hirayama}\ \emph {et~al.}(2018)\citenamefont
  {Hirayama}, \citenamefont {Yamaji}, \citenamefont {Misawa},\ and\
  \citenamefont {Imada}}]{hirayama2018b}%
  \BibitemOpen
  \bibfield  {author} {\bibinfo {author} {\bibfnamefont {M.}~\bibnamefont
  {Hirayama}}, \bibinfo {author} {\bibfnamefont {Y.}~\bibnamefont {Yamaji}},
  \bibinfo {author} {\bibfnamefont {T.}~\bibnamefont {Misawa}},\ and\ \bibinfo
  {author} {\bibfnamefont {M.}~\bibnamefont {Imada}},\ }\href
  {https://doi.org/10.1103/PhysRevB.98.134501} {\bibfield  {journal} {\bibinfo
  {journal} {Phys. Rev. B}\ }\textbf {\bibinfo {volume} {98}},\ \bibinfo
  {pages} {134501} (\bibinfo {year} {2018})}\BibitemShut {NoStop}%
\bibitem [{\citenamefont {Chen}\ \emph {et~al.}(2009)\citenamefont {Chen},
  \citenamefont {Fu}, \citenamefont {Franchini},\ and\ \citenamefont
  {Podloucky}}]{chen2009}%
  \BibitemOpen
  \bibfield  {author} {\bibinfo {author} {\bibfnamefont {X.-Q.}\ \bibnamefont
  {Chen}}, \bibinfo {author} {\bibfnamefont {C.~L.}\ \bibnamefont {Fu}},
  \bibinfo {author} {\bibfnamefont {C.}~\bibnamefont {Franchini}},\ and\
  \bibinfo {author} {\bibfnamefont {R.}~\bibnamefont {Podloucky}},\ }\href
  {https://doi.org/10.1103/PhysRevB.80.094527} {\bibfield  {journal} {\bibinfo
  {journal} {Phys. Rev. B}\ }\textbf {\bibinfo {volume} {80}},\ \bibinfo
  {pages} {094527} (\bibinfo {year} {2009})}\BibitemShut {NoStop}%
\bibitem [{\citenamefont {Peralta}\ \emph {et~al.}(2009)\citenamefont
  {Peralta}, \citenamefont {Puggioni}, \citenamefont {Filippetti},\ and\
  \citenamefont {Fiorentini}}]{peralta2009}%
  \BibitemOpen
  \bibfield  {author} {\bibinfo {author} {\bibfnamefont {G.}~\bibnamefont
  {Peralta}}, \bibinfo {author} {\bibfnamefont {D.}~\bibnamefont {Puggioni}},
  \bibinfo {author} {\bibfnamefont {A.}~\bibnamefont {Filippetti}},\ and\
  \bibinfo {author} {\bibfnamefont {V.}~\bibnamefont {Fiorentini}},\ }\href
  {https://doi.org/10.1103/PhysRevB.80.140408} {\bibfield  {journal} {\bibinfo
  {journal} {Phys. Rev. B}\ }\textbf {\bibinfo {volume} {80}},\ \bibinfo
  {pages} {140408} (\bibinfo {year} {2009})}\BibitemShut {NoStop}%
\bibitem [{\citenamefont {Wang}\ \emph {et~al.}(2014)\citenamefont {Wang},
  \citenamefont {Wei}, \citenamefont {Ding}, \citenamefont {Xing},\ and\
  \citenamefont {Chen}}]{wang2014}%
  \BibitemOpen
  \bibfield  {author} {\bibinfo {author} {\bibfnamefont {F.-F.}\ \bibnamefont
  {Wang}}, \bibinfo {author} {\bibfnamefont {P.-Y.}\ \bibnamefont {Wei}},
  \bibinfo {author} {\bibfnamefont {X.-Y.}\ \bibnamefont {Ding}}, \bibinfo
  {author} {\bibfnamefont {X.-R.}\ \bibnamefont {Xing}},\ and\ \bibinfo
  {author} {\bibfnamefont {X.-Q.}\ \bibnamefont {Chen}},\ }\href
  {https://doi.org/10.1088/0256-307X/31/2/027402} {\bibfield  {journal}
  {\bibinfo  {journal} {Chinese Phys. Lett.}\ }\textbf {\bibinfo {volume}
  {31}},\ \bibinfo {pages} {027402} (\bibinfo {year} {2014})}\BibitemShut
  {NoStop}%
\bibitem [{\citenamefont {Moser}\ \emph {et~al.}(2015)\citenamefont {Moser},
  \citenamefont {Shaik}, \citenamefont {Samal}, \citenamefont {Fatale},
  \citenamefont {Dalla~Piazza}, \citenamefont {Dantz}, \citenamefont
  {Pelliciari}, \citenamefont {{Olalde-Velasco}}, \citenamefont {Schmitt},
  \citenamefont {Koster}, \citenamefont {Mila}, \citenamefont {R{\o}nnow},\
  and\ \citenamefont {Grioni}}]{moser2015}%
  \BibitemOpen
  \bibfield  {author} {\bibinfo {author} {\bibfnamefont {S.}~\bibnamefont
  {Moser}}, \bibinfo {author} {\bibfnamefont {N.~E.}\ \bibnamefont {Shaik}},
  \bibinfo {author} {\bibfnamefont {D.}~\bibnamefont {Samal}}, \bibinfo
  {author} {\bibfnamefont {S.}~\bibnamefont {Fatale}}, \bibinfo {author}
  {\bibfnamefont {B.}~\bibnamefont {Dalla~Piazza}}, \bibinfo {author}
  {\bibfnamefont {M.}~\bibnamefont {Dantz}}, \bibinfo {author} {\bibfnamefont
  {J.}~\bibnamefont {Pelliciari}}, \bibinfo {author} {\bibfnamefont
  {P.}~\bibnamefont {{Olalde-Velasco}}}, \bibinfo {author} {\bibfnamefont
  {T.}~\bibnamefont {Schmitt}}, \bibinfo {author} {\bibfnamefont
  {G.}~\bibnamefont {Koster}}, \bibinfo {author} {\bibfnamefont
  {F.}~\bibnamefont {Mila}}, \bibinfo {author} {\bibfnamefont {H.~M.}\
  \bibnamefont {R{\o}nnow}},\ and\ \bibinfo {author} {\bibfnamefont
  {M.}~\bibnamefont {Grioni}},\ }\href
  {https://doi.org/10.1103/PhysRevB.92.140404} {\bibfield  {journal} {\bibinfo
  {journal} {Phys. Rev. B}\ }\textbf {\bibinfo {volume} {92}},\ \bibinfo
  {pages} {140404} (\bibinfo {year} {2015})}\BibitemShut {NoStop}%
\bibitem [{\citenamefont {Ortiz~Hern{\'a}ndez}\ \emph
  {et~al.}(2021)\citenamefont {Ortiz~Hern{\'a}ndez}, \citenamefont {Salman},
  \citenamefont {Prokscha}, \citenamefont {Suter}, \citenamefont {Mardegan},
  \citenamefont {Moser}, \citenamefont {Zakharova}, \citenamefont
  {Piamonteze},\ and\ \citenamefont {Staub}}]{hernandez2021}%
  \BibitemOpen
  \bibfield  {author} {\bibinfo {author} {\bibfnamefont {N.}~\bibnamefont
  {Ortiz~Hern{\'a}ndez}}, \bibinfo {author} {\bibfnamefont {Z.}~\bibnamefont
  {Salman}}, \bibinfo {author} {\bibfnamefont {T.}~\bibnamefont {Prokscha}},
  \bibinfo {author} {\bibfnamefont {A.}~\bibnamefont {Suter}}, \bibinfo
  {author} {\bibfnamefont {J.~R.~L.}\ \bibnamefont {Mardegan}}, \bibinfo
  {author} {\bibfnamefont {S.}~\bibnamefont {Moser}}, \bibinfo {author}
  {\bibfnamefont {A.}~\bibnamefont {Zakharova}}, \bibinfo {author}
  {\bibfnamefont {C.}~\bibnamefont {Piamonteze}},\ and\ \bibinfo {author}
  {\bibfnamefont {U.}~\bibnamefont {Staub}},\ }\href
  {https://doi.org/10.1103/PhysRevB.103.224429} {\bibfield  {journal} {\bibinfo
   {journal} {Phys. Rev. B}\ }\textbf {\bibinfo {volume} {103}},\ \bibinfo
  {pages} {224429} (\bibinfo {year} {2021})}\BibitemShut {NoStop}%
\bibitem [{\citenamefont {Bramberger}\ \emph {et~al.}(2022)\citenamefont
  {Bramberger}, \citenamefont {{Bacq-Labreuil}}, \citenamefont {Grundner},
  \citenamefont {Biermann}, \citenamefont {Schollw{\"o}ck}, \citenamefont
  {Paeckel},\ and\ \citenamefont {Lenz}}]{bramberger2022}%
  \BibitemOpen
  \bibfield  {author} {\bibinfo {author} {\bibfnamefont {M.}~\bibnamefont
  {Bramberger}}, \bibinfo {author} {\bibfnamefont {B.}~\bibnamefont
  {{Bacq-Labreuil}}}, \bibinfo {author} {\bibfnamefont {M.}~\bibnamefont
  {Grundner}}, \bibinfo {author} {\bibfnamefont {S.}~\bibnamefont {Biermann}},
  \bibinfo {author} {\bibfnamefont {U.}~\bibnamefont {Schollw{\"o}ck}},
  \bibinfo {author} {\bibfnamefont {S.}~\bibnamefont {Paeckel}},\ and\ \bibinfo
  {author} {\bibfnamefont {B.}~\bibnamefont {Lenz}},\ }\href@noop {} {\bibfield
   {journal} {\bibinfo  {journal} {arXiv:2203.07880 [cond-mat]}\ } (\bibinfo
  {year} {2022})},\ \Eprint {https://arxiv.org/abs/2203.07880}
  {arXiv:2203.07880 [cond-mat]} \BibitemShut {NoStop}%
\bibitem [{\citenamefont {Moser}\ \emph {et~al.}(2014)\citenamefont {Moser},
  \citenamefont {Moreschini}, \citenamefont {Yang}, \citenamefont {Innocenti},
  \citenamefont {Fuchs}, \citenamefont {Hansen}, \citenamefont {Chang},
  \citenamefont {Kim}, \citenamefont {Walter}, \citenamefont {Bostwick},
  \citenamefont {Rotenberg}, \citenamefont {Mila},\ and\ \citenamefont
  {Grioni}}]{moser2014}%
  \BibitemOpen
  \bibfield  {author} {\bibinfo {author} {\bibfnamefont {S.}~\bibnamefont
  {Moser}}, \bibinfo {author} {\bibfnamefont {L.}~\bibnamefont {Moreschini}},
  \bibinfo {author} {\bibfnamefont {H.-Y.}\ \bibnamefont {Yang}}, \bibinfo
  {author} {\bibfnamefont {D.}~\bibnamefont {Innocenti}}, \bibinfo {author}
  {\bibfnamefont {F.}~\bibnamefont {Fuchs}}, \bibinfo {author} {\bibfnamefont
  {N.~H.}\ \bibnamefont {Hansen}}, \bibinfo {author} {\bibfnamefont {Y.~J.}\
  \bibnamefont {Chang}}, \bibinfo {author} {\bibfnamefont {K.~S.}\ \bibnamefont
  {Kim}}, \bibinfo {author} {\bibfnamefont {A.~L.}\ \bibnamefont {Walter}},
  \bibinfo {author} {\bibfnamefont {A.}~\bibnamefont {Bostwick}}, \bibinfo
  {author} {\bibfnamefont {E.}~\bibnamefont {Rotenberg}}, \bibinfo {author}
  {\bibfnamefont {F.}~\bibnamefont {Mila}},\ and\ \bibinfo {author}
  {\bibfnamefont {M.}~\bibnamefont {Grioni}},\ }\href
  {https://doi.org/10.1103/PhysRevLett.113.187001} {\bibfield  {journal}
  {\bibinfo  {journal} {Phys. Rev. Lett.}\ }\textbf {\bibinfo {volume} {113}},\
  \bibinfo {pages} {187001} (\bibinfo {year} {2014})}\BibitemShut {NoStop}%
\bibitem [{\citenamefont {Salluzzo}\ \emph {et~al.}(2013)\citenamefont
  {Salluzzo}, \citenamefont {Gariglio}, \citenamefont {Stornaiuolo},
  \citenamefont {Sessi}, \citenamefont {Rusponi}, \citenamefont {Piamonteze},
  \citenamefont {De~Luca}, \citenamefont {Minola}, \citenamefont {Marr{\'e}},
  \citenamefont {Gadaleta}, \citenamefont {Brune}, \citenamefont {Nolting},
  \citenamefont {Brookes},\ and\ \citenamefont {Ghiringhelli}}]{salluzzo2013}%
  \BibitemOpen
  \bibfield  {author} {\bibinfo {author} {\bibfnamefont {M.}~\bibnamefont
  {Salluzzo}}, \bibinfo {author} {\bibfnamefont {S.}~\bibnamefont {Gariglio}},
  \bibinfo {author} {\bibfnamefont {D.}~\bibnamefont {Stornaiuolo}}, \bibinfo
  {author} {\bibfnamefont {V.}~\bibnamefont {Sessi}}, \bibinfo {author}
  {\bibfnamefont {S.}~\bibnamefont {Rusponi}}, \bibinfo {author} {\bibfnamefont
  {C.}~\bibnamefont {Piamonteze}}, \bibinfo {author} {\bibfnamefont {G.~M.}\
  \bibnamefont {De~Luca}}, \bibinfo {author} {\bibfnamefont {M.}~\bibnamefont
  {Minola}}, \bibinfo {author} {\bibfnamefont {D.}~\bibnamefont {Marr{\'e}}},
  \bibinfo {author} {\bibfnamefont {A.}~\bibnamefont {Gadaleta}}, \bibinfo
  {author} {\bibfnamefont {H.}~\bibnamefont {Brune}}, \bibinfo {author}
  {\bibfnamefont {F.}~\bibnamefont {Nolting}}, \bibinfo {author} {\bibfnamefont
  {N.~B.}\ \bibnamefont {Brookes}},\ and\ \bibinfo {author} {\bibfnamefont
  {G.}~\bibnamefont {Ghiringhelli}},\ }\href
  {https://doi.org/10.1103/PhysRevLett.111.087204} {\bibfield  {journal}
  {\bibinfo  {journal} {Phys. Rev. Lett.}\ }\textbf {\bibinfo {volume} {111}},\
  \bibinfo {pages} {087204} (\bibinfo {year} {2013})}\BibitemShut {NoStop}%
\bibitem [{\citenamefont {Lee}\ \emph {et~al.}(2013)\citenamefont {Lee},
  \citenamefont {Xie}, \citenamefont {Sato}, \citenamefont {Bell},
  \citenamefont {Hikita}, \citenamefont {Hwang},\ and\ \citenamefont
  {Kao}}]{lee2013}%
  \BibitemOpen
  \bibfield  {author} {\bibinfo {author} {\bibfnamefont {J.-S.}\ \bibnamefont
  {Lee}}, \bibinfo {author} {\bibfnamefont {Y.~W.}\ \bibnamefont {Xie}},
  \bibinfo {author} {\bibfnamefont {H.~K.}\ \bibnamefont {Sato}}, \bibinfo
  {author} {\bibfnamefont {C.}~\bibnamefont {Bell}}, \bibinfo {author}
  {\bibfnamefont {Y.}~\bibnamefont {Hikita}}, \bibinfo {author} {\bibfnamefont
  {H.~Y.}\ \bibnamefont {Hwang}},\ and\ \bibinfo {author} {\bibfnamefont
  {C.-C.}\ \bibnamefont {Kao}},\ }\href {https://doi.org/10.1038/nmat3674}
  {\bibfield  {journal} {\bibinfo  {journal} {Nature Mater}\ }\textbf {\bibinfo
  {volume} {12}},\ \bibinfo {pages} {703} (\bibinfo {year} {2013})}\BibitemShut
  {NoStop}%
\bibitem [{\citenamefont {Mardegan}\ \emph {et~al.}(2019)\citenamefont
  {Mardegan}, \citenamefont {Christensen}, \citenamefont {Chen}, \citenamefont
  {Parchenko}, \citenamefont {Avula}, \citenamefont {{Ortiz-Hernandez}},
  \citenamefont {Decker}, \citenamefont {Piamonteze}, \citenamefont {Pryds},\
  and\ \citenamefont {Staub}}]{mardegan2019}%
  \BibitemOpen
  \bibfield  {author} {\bibinfo {author} {\bibfnamefont {J.~R.~L.}\
  \bibnamefont {Mardegan}}, \bibinfo {author} {\bibfnamefont {D.~V.}\
  \bibnamefont {Christensen}}, \bibinfo {author} {\bibfnamefont {Y.~Z.}\
  \bibnamefont {Chen}}, \bibinfo {author} {\bibfnamefont {S.}~\bibnamefont
  {Parchenko}}, \bibinfo {author} {\bibfnamefont {S.~R.~V.}\ \bibnamefont
  {Avula}}, \bibinfo {author} {\bibfnamefont {N.}~\bibnamefont
  {{Ortiz-Hernandez}}}, \bibinfo {author} {\bibfnamefont {M.}~\bibnamefont
  {Decker}}, \bibinfo {author} {\bibfnamefont {C.}~\bibnamefont {Piamonteze}},
  \bibinfo {author} {\bibfnamefont {N.}~\bibnamefont {Pryds}},\ and\ \bibinfo
  {author} {\bibfnamefont {U.}~\bibnamefont {Staub}},\ }\href
  {https://doi.org/10.1103/PhysRevB.99.134423} {\bibfield  {journal} {\bibinfo
  {journal} {Phys. Rev. B}\ }\textbf {\bibinfo {volume} {99}},\ \bibinfo
  {pages} {134423} (\bibinfo {year} {2019})}\BibitemShut {NoStop}%
\bibitem [{\citenamefont {R{\"o}del}\ \emph {et~al.}(2018)\citenamefont
  {R{\"o}del}, \citenamefont {Dai}, \citenamefont {Fortuna}, \citenamefont
  {Frantzeskakis}, \citenamefont {Le~F{\`e}vre}, \citenamefont {Bertran},
  \citenamefont {Kobayashi}, \citenamefont {Yukawa}, \citenamefont
  {Mitsuhashi}, \citenamefont {Kitamura}, \citenamefont {Horiba}, \citenamefont
  {Kumigashira},\ and\ \citenamefont {{Santander-Syro}}}]{rodel2018}%
  \BibitemOpen
  \bibfield  {author} {\bibinfo {author} {\bibfnamefont {T.~C.}\ \bibnamefont
  {R{\"o}del}}, \bibinfo {author} {\bibfnamefont {J.}~\bibnamefont {Dai}},
  \bibinfo {author} {\bibfnamefont {F.}~\bibnamefont {Fortuna}}, \bibinfo
  {author} {\bibfnamefont {E.}~\bibnamefont {Frantzeskakis}}, \bibinfo {author}
  {\bibfnamefont {P.}~\bibnamefont {Le~F{\`e}vre}}, \bibinfo {author}
  {\bibfnamefont {F.}~\bibnamefont {Bertran}}, \bibinfo {author} {\bibfnamefont
  {M.}~\bibnamefont {Kobayashi}}, \bibinfo {author} {\bibfnamefont
  {R.}~\bibnamefont {Yukawa}}, \bibinfo {author} {\bibfnamefont
  {T.}~\bibnamefont {Mitsuhashi}}, \bibinfo {author} {\bibfnamefont
  {M.}~\bibnamefont {Kitamura}}, \bibinfo {author} {\bibfnamefont
  {K.}~\bibnamefont {Horiba}}, \bibinfo {author} {\bibfnamefont
  {H.}~\bibnamefont {Kumigashira}},\ and\ \bibinfo {author} {\bibfnamefont
  {A.~F.}\ \bibnamefont {{Santander-Syro}}},\ }\href
  {https://doi.org/10.1103/PhysRevMaterials.2.051601} {\bibfield  {journal}
  {\bibinfo  {journal} {Phys. Rev. Materials}\ }\textbf {\bibinfo {volume}
  {2}},\ \bibinfo {pages} {051601} (\bibinfo {year} {2018})}\BibitemShut
  {NoStop}%
\bibitem [{\citenamefont {{Santander-Syro}}\ \emph {et~al.}(2011)\citenamefont
  {{Santander-Syro}}, \citenamefont {Copie}, \citenamefont {Kondo},
  \citenamefont {Fortuna}, \citenamefont {Pailh{\`e}s}, \citenamefont {Weht},
  \citenamefont {Qiu}, \citenamefont {Bertran}, \citenamefont {Nicolaou},
  \citenamefont {{Taleb-Ibrahimi}}, \citenamefont {Le~F{\`e}vre}, \citenamefont
  {Herranz}, \citenamefont {Bibes}, \citenamefont {Reyren}, \citenamefont
  {Apertet}, \citenamefont {Lecoeur}, \citenamefont {Barth{\'e}l{\'e}my},\ and\
  \citenamefont {Rozenberg}}]{santander2011}%
  \BibitemOpen
  \bibfield  {author} {\bibinfo {author} {\bibfnamefont {A.~F.}\ \bibnamefont
  {{Santander-Syro}}}, \bibinfo {author} {\bibfnamefont {O.}~\bibnamefont
  {Copie}}, \bibinfo {author} {\bibfnamefont {T.}~\bibnamefont {Kondo}},
  \bibinfo {author} {\bibfnamefont {F.}~\bibnamefont {Fortuna}}, \bibinfo
  {author} {\bibfnamefont {S.}~\bibnamefont {Pailh{\`e}s}}, \bibinfo {author}
  {\bibfnamefont {R.}~\bibnamefont {Weht}}, \bibinfo {author} {\bibfnamefont
  {X.~G.}\ \bibnamefont {Qiu}}, \bibinfo {author} {\bibfnamefont
  {F.}~\bibnamefont {Bertran}}, \bibinfo {author} {\bibfnamefont
  {A.}~\bibnamefont {Nicolaou}}, \bibinfo {author} {\bibfnamefont
  {A.}~\bibnamefont {{Taleb-Ibrahimi}}}, \bibinfo {author} {\bibfnamefont
  {P.}~\bibnamefont {Le~F{\`e}vre}}, \bibinfo {author} {\bibfnamefont
  {G.}~\bibnamefont {Herranz}}, \bibinfo {author} {\bibfnamefont
  {M.}~\bibnamefont {Bibes}}, \bibinfo {author} {\bibfnamefont
  {N.}~\bibnamefont {Reyren}}, \bibinfo {author} {\bibfnamefont
  {Y.}~\bibnamefont {Apertet}}, \bibinfo {author} {\bibfnamefont
  {P.}~\bibnamefont {Lecoeur}}, \bibinfo {author} {\bibfnamefont
  {A.}~\bibnamefont {Barth{\'e}l{\'e}my}},\ and\ \bibinfo {author}
  {\bibfnamefont {M.~J.}\ \bibnamefont {Rozenberg}},\ }\href
  {https://doi.org/10.1038/nature09720} {\bibfield  {journal} {\bibinfo
  {journal} {Nature}\ }\textbf {\bibinfo {volume} {469}},\ \bibinfo {pages}
  {189} (\bibinfo {year} {2011})}\BibitemShut {NoStop}%
\bibitem [{\citenamefont {Chen}\ \emph {et~al.}(2013)\citenamefont {Chen},
  \citenamefont {Bovet}, \citenamefont {Trier}, \citenamefont {Christensen},
  \citenamefont {Qu}, \citenamefont {Andersen}, \citenamefont {Kasama},
  \citenamefont {Zhang}, \citenamefont {Giraud}, \citenamefont {Dufouleur},
  \citenamefont {Jespersen}, \citenamefont {Sun}, \citenamefont {Smith},
  \citenamefont {Nyg{\aa}rd}, \citenamefont {Lu}, \citenamefont {B{\"u}chner},
  \citenamefont {Shen}, \citenamefont {Linderoth},\ and\ \citenamefont
  {Pryds}}]{chen2013a}%
  \BibitemOpen
  \bibfield  {author} {\bibinfo {author} {\bibfnamefont {Y.~Z.}\ \bibnamefont
  {Chen}}, \bibinfo {author} {\bibfnamefont {N.}~\bibnamefont {Bovet}},
  \bibinfo {author} {\bibfnamefont {F.}~\bibnamefont {Trier}}, \bibinfo
  {author} {\bibfnamefont {D.~V.}\ \bibnamefont {Christensen}}, \bibinfo
  {author} {\bibfnamefont {F.~M.}\ \bibnamefont {Qu}}, \bibinfo {author}
  {\bibfnamefont {N.~H.}\ \bibnamefont {Andersen}}, \bibinfo {author}
  {\bibfnamefont {T.}~\bibnamefont {Kasama}}, \bibinfo {author} {\bibfnamefont
  {W.}~\bibnamefont {Zhang}}, \bibinfo {author} {\bibfnamefont
  {R.}~\bibnamefont {Giraud}}, \bibinfo {author} {\bibfnamefont
  {J.}~\bibnamefont {Dufouleur}}, \bibinfo {author} {\bibfnamefont {T.~S.}\
  \bibnamefont {Jespersen}}, \bibinfo {author} {\bibfnamefont {J.~R.}\
  \bibnamefont {Sun}}, \bibinfo {author} {\bibfnamefont {A.}~\bibnamefont
  {Smith}}, \bibinfo {author} {\bibfnamefont {J.}~\bibnamefont {Nyg{\aa}rd}},
  \bibinfo {author} {\bibfnamefont {L.}~\bibnamefont {Lu}}, \bibinfo {author}
  {\bibfnamefont {B.}~\bibnamefont {B{\"u}chner}}, \bibinfo {author}
  {\bibfnamefont {B.~G.}\ \bibnamefont {Shen}}, \bibinfo {author}
  {\bibfnamefont {S.}~\bibnamefont {Linderoth}},\ and\ \bibinfo {author}
  {\bibfnamefont {N.}~\bibnamefont {Pryds}},\ }\href
  {https://doi.org/10.1038/ncomms2394} {\bibfield  {journal} {\bibinfo
  {journal} {Nat Commun}\ }\textbf {\bibinfo {volume} {4}},\ \bibinfo {pages}
  {1371} (\bibinfo {year} {2013})}\BibitemShut {NoStop}%
\bibitem [{\citenamefont {Taniuchi}\ \emph {et~al.}(2016)\citenamefont
  {Taniuchi}, \citenamefont {Motoyui}, \citenamefont {Morozumi}, \citenamefont
  {R{\"o}del}, \citenamefont {Fortuna}, \citenamefont {{Santander-Syro}},\ and\
  \citenamefont {Shin}}]{taniuchi2016}%
  \BibitemOpen
  \bibfield  {author} {\bibinfo {author} {\bibfnamefont {T.}~\bibnamefont
  {Taniuchi}}, \bibinfo {author} {\bibfnamefont {Y.}~\bibnamefont {Motoyui}},
  \bibinfo {author} {\bibfnamefont {K.}~\bibnamefont {Morozumi}}, \bibinfo
  {author} {\bibfnamefont {T.~C.}\ \bibnamefont {R{\"o}del}}, \bibinfo {author}
  {\bibfnamefont {F.}~\bibnamefont {Fortuna}}, \bibinfo {author} {\bibfnamefont
  {A.~F.}\ \bibnamefont {{Santander-Syro}}},\ and\ \bibinfo {author}
  {\bibfnamefont {S.}~\bibnamefont {Shin}},\ }\href
  {https://doi.org/10.1038/ncomms11781} {\bibfield  {journal} {\bibinfo
  {journal} {Nat Commun}\ }\textbf {\bibinfo {volume} {7}},\ \bibinfo {pages}
  {11781} (\bibinfo {year} {2016})}\BibitemShut {NoStop}%
\bibitem [{\citenamefont {Altmeyer}\ \emph {et~al.}(2016)\citenamefont
  {Altmeyer}, \citenamefont {Jeschke}, \citenamefont {{Hijano-Cubelos}},
  \citenamefont {Martins}, \citenamefont {Lechermann}, \citenamefont
  {Koepernik}, \citenamefont {{Santander-Syro}}, \citenamefont {Rozenberg},
  \citenamefont {Valent{\'i}},\ and\ \citenamefont {Gabay}}]{altmeyer2016}%
  \BibitemOpen
  \bibfield  {author} {\bibinfo {author} {\bibfnamefont {M.}~\bibnamefont
  {Altmeyer}}, \bibinfo {author} {\bibfnamefont {H.~O.}\ \bibnamefont
  {Jeschke}}, \bibinfo {author} {\bibfnamefont {O.}~\bibnamefont
  {{Hijano-Cubelos}}}, \bibinfo {author} {\bibfnamefont {C.}~\bibnamefont
  {Martins}}, \bibinfo {author} {\bibfnamefont {F.}~\bibnamefont {Lechermann}},
  \bibinfo {author} {\bibfnamefont {K.}~\bibnamefont {Koepernik}}, \bibinfo
  {author} {\bibfnamefont {A.~F.}\ \bibnamefont {{Santander-Syro}}}, \bibinfo
  {author} {\bibfnamefont {M.~J.}\ \bibnamefont {Rozenberg}}, \bibinfo {author}
  {\bibfnamefont {R.}~\bibnamefont {Valent{\'i}}},\ and\ \bibinfo {author}
  {\bibfnamefont {M.}~\bibnamefont {Gabay}},\ }\href
  {https://doi.org/10.1103/PhysRevLett.116.157203} {\bibfield  {journal}
  {\bibinfo  {journal} {Phys. Rev. Lett.}\ }\textbf {\bibinfo {volume} {116}},\
  \bibinfo {pages} {157203} (\bibinfo {year} {2016})}\BibitemShut {NoStop}%
\bibitem [{\citenamefont {Franchini}\ \emph {et~al.}(2011)\citenamefont
  {Franchini}, \citenamefont {Chen},\ and\ \citenamefont
  {Podloucky}}]{franchini2011}%
  \BibitemOpen
  \bibfield  {author} {\bibinfo {author} {\bibfnamefont {C.}~\bibnamefont
  {Franchini}}, \bibinfo {author} {\bibfnamefont {X.-Q.}\ \bibnamefont
  {Chen}},\ and\ \bibinfo {author} {\bibfnamefont {R.}~\bibnamefont
  {Podloucky}},\ }\href {https://doi.org/10.1088/0953-8984/23/4/045004}
  {\bibfield  {journal} {\bibinfo  {journal} {J. Phys.: Condens. Matter}\
  }\textbf {\bibinfo {volume} {23}},\ \bibinfo {pages} {045004} (\bibinfo
  {year} {2011})}\BibitemShut {NoStop}%
\bibitem [{\citenamefont {Drera}\ \emph {et~al.}(2019)\citenamefont {Drera},
  \citenamefont {Giampietri}, \citenamefont {Febbrari}, \citenamefont
  {Patrini}, \citenamefont {Mozzati},\ and\ \citenamefont
  {Sangaletti}}]{drera2019}%
  \BibitemOpen
  \bibfield  {author} {\bibinfo {author} {\bibfnamefont {G.}~\bibnamefont
  {Drera}}, \bibinfo {author} {\bibfnamefont {A.}~\bibnamefont {Giampietri}},
  \bibinfo {author} {\bibfnamefont {A.}~\bibnamefont {Febbrari}}, \bibinfo
  {author} {\bibfnamefont {M.}~\bibnamefont {Patrini}}, \bibinfo {author}
  {\bibfnamefont {M.~C.}\ \bibnamefont {Mozzati}},\ and\ \bibinfo {author}
  {\bibfnamefont {L.}~\bibnamefont {Sangaletti}},\ }\href
  {https://doi.org/10.1103/PhysRevB.99.075124} {\bibfield  {journal} {\bibinfo
  {journal} {Phys. Rev. B}\ }\textbf {\bibinfo {volume} {99}},\ \bibinfo
  {pages} {075124} (\bibinfo {year} {2019})}\BibitemShut {NoStop}%
\bibitem [{\citenamefont {Liechtenstein}\ \emph {et~al.}(1995)\citenamefont
  {Liechtenstein}, \citenamefont {Anisimov},\ and\ \citenamefont
  {Zaanen}}]{liechtenstein1995}%
  \BibitemOpen
  \bibfield  {author} {\bibinfo {author} {\bibfnamefont {A.~I.}\ \bibnamefont
  {Liechtenstein}}, \bibinfo {author} {\bibfnamefont {V.~I.}\ \bibnamefont
  {Anisimov}},\ and\ \bibinfo {author} {\bibfnamefont {J.}~\bibnamefont
  {Zaanen}},\ }\href {https://doi.org/10.1103/PhysRevB.52.R5467} {\bibfield
  {journal} {\bibinfo  {journal} {Phys. Rev. B}\ }\textbf {\bibinfo {volume}
  {52}},\ \bibinfo {pages} {R5467} (\bibinfo {year} {1995})}\BibitemShut
  {NoStop}%
\bibitem [{\citenamefont {Perdew}\ \emph {et~al.}(1996)\citenamefont {Perdew},
  \citenamefont {Burke},\ and\ \citenamefont {Ernzerhof}}]{perdew1996}%
  \BibitemOpen
  \bibfield  {author} {\bibinfo {author} {\bibfnamefont {J.~P.}\ \bibnamefont
  {Perdew}}, \bibinfo {author} {\bibfnamefont {K.}~\bibnamefont {Burke}},\ and\
  \bibinfo {author} {\bibfnamefont {M.}~\bibnamefont {Ernzerhof}},\ }\href
  {https://doi.org/10.1103/PhysRevLett.77.3865} {\bibfield  {journal} {\bibinfo
   {journal} {Phys. Rev. Lett.}\ }\textbf {\bibinfo {volume} {77}},\ \bibinfo
  {pages} {3865} (\bibinfo {year} {1996})}\BibitemShut {NoStop}%
\bibitem [{\citenamefont {Perdew}\ \emph {et~al.}(1997)\citenamefont {Perdew},
  \citenamefont {Burke},\ and\ \citenamefont {Ernzerhof}}]{perdew1997}%
  \BibitemOpen
  \bibfield  {author} {\bibinfo {author} {\bibfnamefont {J.~P.}\ \bibnamefont
  {Perdew}}, \bibinfo {author} {\bibfnamefont {K.}~\bibnamefont {Burke}},\ and\
  \bibinfo {author} {\bibfnamefont {M.}~\bibnamefont {Ernzerhof}},\ }\href
  {https://doi.org/10.1103/PhysRevLett.78.1396} {\bibfield  {journal} {\bibinfo
   {journal} {Phys. Rev. Lett.}\ }\textbf {\bibinfo {volume} {78}},\ \bibinfo
  {pages} {1396} (\bibinfo {year} {1997})}\BibitemShut {NoStop}%
\bibitem [{\citenamefont {Wahl}\ \emph {et~al.}(2008)\citenamefont {Wahl},
  \citenamefont {Vogtenhuber},\ and\ \citenamefont {Kresse}}]{wahl2008}%
  \BibitemOpen
  \bibfield  {author} {\bibinfo {author} {\bibfnamefont {R.}~\bibnamefont
  {Wahl}}, \bibinfo {author} {\bibfnamefont {D.}~\bibnamefont {Vogtenhuber}},\
  and\ \bibinfo {author} {\bibfnamefont {G.}~\bibnamefont {Kresse}},\ }\href
  {https://doi.org/10.1103/PhysRevB.78.104116} {\bibfield  {journal} {\bibinfo
  {journal} {Phys. Rev. B}\ }\textbf {\bibinfo {volume} {78}},\ \bibinfo
  {pages} {104116} (\bibinfo {year} {2008})}\BibitemShut {NoStop}%
\bibitem [{\citenamefont {Becke}(1993)}]{becke1993}%
  \BibitemOpen
  \bibfield  {author} {\bibinfo {author} {\bibfnamefont {A.~D.}\ \bibnamefont
  {Becke}},\ }\href {https://doi.org/10.1063/1.464304} {\bibfield  {journal}
  {\bibinfo  {journal} {The Journal of Chemical Physics}\ }\textbf {\bibinfo
  {volume} {98}},\ \bibinfo {pages} {1372} (\bibinfo {year}
  {1993})}\BibitemShut {NoStop}%
\bibitem [{\citenamefont {Cao}\ \emph {et~al.}(2000)\citenamefont {Cao},
  \citenamefont {Sozontov},\ and\ \citenamefont {Zegenhagen}}]{cao2000}%
  \BibitemOpen
  \bibfield  {author} {\bibinfo {author} {\bibfnamefont {L.}~\bibnamefont
  {Cao}}, \bibinfo {author} {\bibfnamefont {E.}~\bibnamefont {Sozontov}},\ and\
  \bibinfo {author} {\bibfnamefont {J.}~\bibnamefont {Zegenhagen}},\ }\href
  {https://doi.org/10.1002/1521-396X(200010)181:2<387::AID-PSSA387>3.0.CO;2-5}
  {\bibfield  {journal} {\bibinfo  {journal} {physica status solidi (a)}\
  }\textbf {\bibinfo {volume} {181}},\ \bibinfo {pages} {387} (\bibinfo {year}
  {2000})}\BibitemShut {NoStop}%
\bibitem [{\citenamefont {Kresse}\ and\ \citenamefont
  {Hafner}(1993)}]{kresse1993}%
  \BibitemOpen
  \bibfield  {author} {\bibinfo {author} {\bibfnamefont {G.}~\bibnamefont
  {Kresse}}\ and\ \bibinfo {author} {\bibfnamefont {J.}~\bibnamefont
  {Hafner}},\ }\href {https://doi.org/10.1103/PhysRevB.47.558} {\bibfield
  {journal} {\bibinfo  {journal} {Phys. Rev. B}\ }\textbf {\bibinfo {volume}
  {47}},\ \bibinfo {pages} {558} (\bibinfo {year} {1993})}\BibitemShut
  {NoStop}%
\bibitem [{\citenamefont {Kresse}\ and\ \citenamefont
  {Hafner}(1994)}]{kresse1994}%
  \BibitemOpen
  \bibfield  {author} {\bibinfo {author} {\bibfnamefont {G.}~\bibnamefont
  {Kresse}}\ and\ \bibinfo {author} {\bibfnamefont {J.}~\bibnamefont
  {Hafner}},\ }\href {https://doi.org/10.1103/PhysRevB.49.14251} {\bibfield
  {journal} {\bibinfo  {journal} {Phys. Rev. B}\ }\textbf {\bibinfo {volume}
  {49}},\ \bibinfo {pages} {14251} (\bibinfo {year} {1994})}\BibitemShut
  {NoStop}%
\bibitem [{\citenamefont {Kresse}\ and\ \citenamefont
  {Furthm{\"u}ller}(1996{\natexlab{a}})}]{kresse1996a}%
  \BibitemOpen
  \bibfield  {author} {\bibinfo {author} {\bibfnamefont {G.}~\bibnamefont
  {Kresse}}\ and\ \bibinfo {author} {\bibfnamefont {J.}~\bibnamefont
  {Furthm{\"u}ller}},\ }\href {https://doi.org/10.1016/0927-0256(96)00008-0}
  {\bibfield  {journal} {\bibinfo  {journal} {Computational Materials Science}\
  }\textbf {\bibinfo {volume} {6}},\ \bibinfo {pages} {15} (\bibinfo {year}
  {1996}{\natexlab{a}})}\BibitemShut {NoStop}%
\bibitem [{\citenamefont {Kresse}\ and\ \citenamefont
  {Furthm{\"u}ller}(1996{\natexlab{b}})}]{kresse1996b}%
  \BibitemOpen
  \bibfield  {author} {\bibinfo {author} {\bibfnamefont {G.}~\bibnamefont
  {Kresse}}\ and\ \bibinfo {author} {\bibfnamefont {J.}~\bibnamefont
  {Furthm{\"u}ller}},\ }\href {https://doi.org/10.1103/PhysRevB.54.11169}
  {\bibfield  {journal} {\bibinfo  {journal} {Phys. Rev. B}\ }\textbf {\bibinfo
  {volume} {54}},\ \bibinfo {pages} {11169} (\bibinfo {year}
  {1996}{\natexlab{b}})}\BibitemShut {NoStop}%
\bibitem [{\citenamefont {Kresse}\ and\ \citenamefont
  {Joubert}(1999)}]{kresse1999}%
  \BibitemOpen
  \bibfield  {author} {\bibinfo {author} {\bibfnamefont {G.}~\bibnamefont
  {Kresse}}\ and\ \bibinfo {author} {\bibfnamefont {D.}~\bibnamefont
  {Joubert}},\ }\href {https://doi.org/10.1103/PhysRevB.59.1758} {\bibfield
  {journal} {\bibinfo  {journal} {Phys. Rev. B}\ }\textbf {\bibinfo {volume}
  {59}},\ \bibinfo {pages} {1758} (\bibinfo {year} {1999})}\BibitemShut
  {NoStop}%
\bibitem [{\citenamefont {Schwarz}()}]{wien2k}%
  \BibitemOpen
  \bibfield  {author} {\bibinfo {author} {\bibfnamefont {D.~K.}\ \bibnamefont
  {Schwarz}},\ }\href@noop {} {\ ,\ \bibinfo {pages} {295}}\BibitemShut
  {NoStop}%
\bibitem [{\citenamefont {Blaha}\ \emph {et~al.}(2020)\citenamefont {Blaha},
  \citenamefont {Schwarz}, \citenamefont {Tran}, \citenamefont {Laskowski},
  \citenamefont {Madsen},\ and\ \citenamefont {Marks}}]{blaha2020}%
  \BibitemOpen
  \bibfield  {author} {\bibinfo {author} {\bibfnamefont {P.}~\bibnamefont
  {Blaha}}, \bibinfo {author} {\bibfnamefont {K.}~\bibnamefont {Schwarz}},
  \bibinfo {author} {\bibfnamefont {F.}~\bibnamefont {Tran}}, \bibinfo {author}
  {\bibfnamefont {R.}~\bibnamefont {Laskowski}}, \bibinfo {author}
  {\bibfnamefont {G.~K.~H.}\ \bibnamefont {Madsen}},\ and\ \bibinfo {author}
  {\bibfnamefont {L.~D.}\ \bibnamefont {Marks}},\ }\href
  {https://doi.org/10.1063/1.5143061} {\bibfield  {journal} {\bibinfo
  {journal} {J. Chem. Phys.}\ }\textbf {\bibinfo {volume} {152}},\ \bibinfo
  {pages} {074101} (\bibinfo {year} {2020})}\BibitemShut {NoStop}%
\bibitem [{\citenamefont {Mostofi}\ \emph {et~al.}(2008)\citenamefont
  {Mostofi}, \citenamefont {Yates}, \citenamefont {Lee}, \citenamefont {Souza},
  \citenamefont {Vanderbilt},\ and\ \citenamefont {Marzari}}]{mostofi2008}%
  \BibitemOpen
  \bibfield  {author} {\bibinfo {author} {\bibfnamefont {A.~A.}\ \bibnamefont
  {Mostofi}}, \bibinfo {author} {\bibfnamefont {J.~R.}\ \bibnamefont {Yates}},
  \bibinfo {author} {\bibfnamefont {Y.-S.}\ \bibnamefont {Lee}}, \bibinfo
  {author} {\bibfnamefont {I.}~\bibnamefont {Souza}}, \bibinfo {author}
  {\bibfnamefont {D.}~\bibnamefont {Vanderbilt}},\ and\ \bibinfo {author}
  {\bibfnamefont {N.}~\bibnamefont {Marzari}},\ }\href
  {https://doi.org/10.1016/j.cpc.2007.11.016} {\bibfield  {journal} {\bibinfo
  {journal} {Computer Physics Communications}\ }\textbf {\bibinfo {volume}
  {178}},\ \bibinfo {pages} {685} (\bibinfo {year} {2008})}\BibitemShut
  {NoStop}%
\bibitem [{\citenamefont {Padilla}\ and\ \citenamefont
  {Vanderbilt}(1997)}]{padilla1997}%
  \BibitemOpen
  \bibfield  {author} {\bibinfo {author} {\bibfnamefont {J.}~\bibnamefont
  {Padilla}}\ and\ \bibinfo {author} {\bibfnamefont {D.}~\bibnamefont
  {Vanderbilt}},\ }\href {https://doi.org/10.1103/PhysRevB.56.1625} {\bibfield
  {journal} {\bibinfo  {journal} {Phys. Rev. B}\ }\textbf {\bibinfo {volume}
  {56}},\ \bibinfo {pages} {1625} (\bibinfo {year} {1997})}\BibitemShut
  {NoStop}%
\bibitem [{\citenamefont {Padilla}\ and\ \citenamefont
  {Vanderbilt}(1998)}]{padilla1998}%
  \BibitemOpen
  \bibfield  {author} {\bibinfo {author} {\bibfnamefont {J.}~\bibnamefont
  {Padilla}}\ and\ \bibinfo {author} {\bibfnamefont {D.}~\bibnamefont
  {Vanderbilt}},\ }\href {https://doi.org/10.1016/S0039-6028(98)00670-0}
  {\bibfield  {journal} {\bibinfo  {journal} {Surface Science}\ }\textbf
  {\bibinfo {volume} {418}},\ \bibinfo {pages} {64} (\bibinfo {year}
  {1998})}\BibitemShut {NoStop}%
\bibitem [{\citenamefont {Audehm}\ \emph {et~al.}(2016)\citenamefont {Audehm},
  \citenamefont {Schmidt}, \citenamefont {Br{\"u}ck}, \citenamefont {Tietze},
  \citenamefont {Gr{\"a}fe}, \citenamefont {Macke}, \citenamefont
  {Sch{\"u}tz},\ and\ \citenamefont {Goering}}]{audehm2016}%
  \BibitemOpen
  \bibfield  {author} {\bibinfo {author} {\bibfnamefont {P.}~\bibnamefont
  {Audehm}}, \bibinfo {author} {\bibfnamefont {M.}~\bibnamefont {Schmidt}},
  \bibinfo {author} {\bibfnamefont {S.}~\bibnamefont {Br{\"u}ck}}, \bibinfo
  {author} {\bibfnamefont {T.}~\bibnamefont {Tietze}}, \bibinfo {author}
  {\bibfnamefont {J.}~\bibnamefont {Gr{\"a}fe}}, \bibinfo {author}
  {\bibfnamefont {S.}~\bibnamefont {Macke}}, \bibinfo {author} {\bibfnamefont
  {G.}~\bibnamefont {Sch{\"u}tz}},\ and\ \bibinfo {author} {\bibfnamefont
  {E.}~\bibnamefont {Goering}},\ }\href {https://doi.org/10.1038/srep25517}
  {\bibfield  {journal} {\bibinfo  {journal} {Sci Rep}\ }\textbf {\bibinfo
  {volume} {6}},\ \bibinfo {pages} {25517} (\bibinfo {year}
  {2016})}\BibitemShut {NoStop}%
\bibitem [{\citenamefont {D{\"u}rr}\ and\ \citenamefont {{van der
  Laan}}(1996)}]{durr1996}%
  \BibitemOpen
  \bibfield  {author} {\bibinfo {author} {\bibfnamefont {H.~A.}\ \bibnamefont
  {D{\"u}rr}}\ and\ \bibinfo {author} {\bibfnamefont {G.}~\bibnamefont {{van
  der Laan}}},\ }\href {https://doi.org/10.1103/PhysRevB.54.R760} {\bibfield
  {journal} {\bibinfo  {journal} {Phys. Rev. B}\ }\textbf {\bibinfo {volume}
  {54}},\ \bibinfo {pages} {R760} (\bibinfo {year} {1996})}\BibitemShut
  {NoStop}%
\bibitem [{\citenamefont {Lenz}\ \emph {et~al.}(2019)\citenamefont {Lenz},
  \citenamefont {Martins},\ and\ \citenamefont {Biermann}}]{lenz2019}%
  \BibitemOpen
  \bibfield  {author} {\bibinfo {author} {\bibfnamefont {B.}~\bibnamefont
  {Lenz}}, \bibinfo {author} {\bibfnamefont {C.}~\bibnamefont {Martins}},\ and\
  \bibinfo {author} {\bibfnamefont {S.}~\bibnamefont {Biermann}},\ }\href
  {https://doi.org/10.1088/1361-648X/ab146a} {\bibfield  {journal} {\bibinfo
  {journal} {J. Phys.: Condens. Matter}\ }\textbf {\bibinfo {volume} {31}},\
  \bibinfo {pages} {293001} (\bibinfo {year} {2019})}\BibitemShut {NoStop}%
\bibitem [{\citenamefont {Backes}\ \emph {et~al.}(2016)\citenamefont {Backes},
  \citenamefont {R{\"o}del}, \citenamefont {Fortuna}, \citenamefont
  {Frantzeskakis}, \citenamefont {Le~F{\`e}vre}, \citenamefont {Bertran},
  \citenamefont {Kobayashi}, \citenamefont {Yukawa}, \citenamefont
  {Mitsuhashi}, \citenamefont {Kitamura}, \citenamefont {Horiba}, \citenamefont
  {Kumigashira}, \citenamefont {{Saint-Martin}}, \citenamefont {Fouchet},
  \citenamefont {Berini}, \citenamefont {Dumont}, \citenamefont {Kim},
  \citenamefont {Lechermann}, \citenamefont {Jeschke}, \citenamefont
  {Rozenberg}, \citenamefont {Valent{\'i}},\ and\ \citenamefont
  {{Santander-Syro}}}]{backes2016}%
  \BibitemOpen
  \bibfield  {author} {\bibinfo {author} {\bibfnamefont {S.}~\bibnamefont
  {Backes}}, \bibinfo {author} {\bibfnamefont {T.~C.}\ \bibnamefont
  {R{\"o}del}}, \bibinfo {author} {\bibfnamefont {F.}~\bibnamefont {Fortuna}},
  \bibinfo {author} {\bibfnamefont {E.}~\bibnamefont {Frantzeskakis}}, \bibinfo
  {author} {\bibfnamefont {P.}~\bibnamefont {Le~F{\`e}vre}}, \bibinfo {author}
  {\bibfnamefont {F.}~\bibnamefont {Bertran}}, \bibinfo {author} {\bibfnamefont
  {M.}~\bibnamefont {Kobayashi}}, \bibinfo {author} {\bibfnamefont
  {R.}~\bibnamefont {Yukawa}}, \bibinfo {author} {\bibfnamefont
  {T.}~\bibnamefont {Mitsuhashi}}, \bibinfo {author} {\bibfnamefont
  {M.}~\bibnamefont {Kitamura}}, \bibinfo {author} {\bibfnamefont
  {K.}~\bibnamefont {Horiba}}, \bibinfo {author} {\bibfnamefont
  {H.}~\bibnamefont {Kumigashira}}, \bibinfo {author} {\bibfnamefont
  {R.}~\bibnamefont {{Saint-Martin}}}, \bibinfo {author} {\bibfnamefont
  {A.}~\bibnamefont {Fouchet}}, \bibinfo {author} {\bibfnamefont
  {B.}~\bibnamefont {Berini}}, \bibinfo {author} {\bibfnamefont
  {Y.}~\bibnamefont {Dumont}}, \bibinfo {author} {\bibfnamefont {A.~J.}\
  \bibnamefont {Kim}}, \bibinfo {author} {\bibfnamefont {F.}~\bibnamefont
  {Lechermann}}, \bibinfo {author} {\bibfnamefont {H.~O.}\ \bibnamefont
  {Jeschke}}, \bibinfo {author} {\bibfnamefont {M.~J.}\ \bibnamefont
  {Rozenberg}}, \bibinfo {author} {\bibfnamefont {R.}~\bibnamefont
  {Valent{\'i}}},\ and\ \bibinfo {author} {\bibfnamefont {A.~F.}\ \bibnamefont
  {{Santander-Syro}}},\ }\href {https://doi.org/10.1103/PhysRevB.94.241110}
  {\bibfield  {journal} {\bibinfo  {journal} {Phys. Rev. B}\ }\textbf {\bibinfo
  {volume} {94}},\ \bibinfo {pages} {241110} (\bibinfo {year}
  {2016})}\BibitemShut {NoStop}%
\bibitem [{\citenamefont {Sing}\ \emph {et~al.}(2009)\citenamefont {Sing},
  \citenamefont {Berner}, \citenamefont {Go{\ss}}, \citenamefont {M{\"u}ller},
  \citenamefont {Ruff}, \citenamefont {Wetscherek}, \citenamefont {Thiel},
  \citenamefont {Mannhart}, \citenamefont {Pauli}, \citenamefont {Schneider},
  \citenamefont {Willmott}, \citenamefont {Gorgoi}, \citenamefont
  {Sch{\"a}fers},\ and\ \citenamefont {Claessen}}]{sing2009}%
  \BibitemOpen
  \bibfield  {author} {\bibinfo {author} {\bibfnamefont {M.}~\bibnamefont
  {Sing}}, \bibinfo {author} {\bibfnamefont {G.}~\bibnamefont {Berner}},
  \bibinfo {author} {\bibfnamefont {K.}~\bibnamefont {Go{\ss}}}, \bibinfo
  {author} {\bibfnamefont {A.}~\bibnamefont {M{\"u}ller}}, \bibinfo {author}
  {\bibfnamefont {A.}~\bibnamefont {Ruff}}, \bibinfo {author} {\bibfnamefont
  {A.}~\bibnamefont {Wetscherek}}, \bibinfo {author} {\bibfnamefont
  {S.}~\bibnamefont {Thiel}}, \bibinfo {author} {\bibfnamefont
  {J.}~\bibnamefont {Mannhart}}, \bibinfo {author} {\bibfnamefont {S.~A.}\
  \bibnamefont {Pauli}}, \bibinfo {author} {\bibfnamefont {C.~W.}\ \bibnamefont
  {Schneider}}, \bibinfo {author} {\bibfnamefont {P.~R.}\ \bibnamefont
  {Willmott}}, \bibinfo {author} {\bibfnamefont {M.}~\bibnamefont {Gorgoi}},
  \bibinfo {author} {\bibfnamefont {F.}~\bibnamefont {Sch{\"a}fers}},\ and\
  \bibinfo {author} {\bibfnamefont {R.}~\bibnamefont {Claessen}},\ }\href
  {https://doi.org/10.1103/PhysRevLett.102.176805} {\bibfield  {journal}
  {\bibinfo  {journal} {Phys. Rev. Lett.}\ }\textbf {\bibinfo {volume} {102}},\
  \bibinfo {pages} {176805} (\bibinfo {year} {2009})}\BibitemShut {NoStop}%
\bibitem [{\citenamefont {Mannhart}\ \emph {et~al.}(2008)\citenamefont
  {Mannhart}, \citenamefont {Blank}, \citenamefont {Hwang}, \citenamefont
  {Millis},\ and\ \citenamefont {Triscone}}]{mannhart2008}%
  \BibitemOpen
  \bibfield  {author} {\bibinfo {author} {\bibfnamefont {J.}~\bibnamefont
  {Mannhart}}, \bibinfo {author} {\bibfnamefont {D.}~\bibnamefont {Blank}},
  \bibinfo {author} {\bibfnamefont {H.}~\bibnamefont {Hwang}}, \bibinfo
  {author} {\bibfnamefont {A.}~\bibnamefont {Millis}},\ and\ \bibinfo {author}
  {\bibfnamefont {J.-M.}\ \bibnamefont {Triscone}},\ }\href
  {https://doi.org/10.1557/mrs2008.222} {\bibfield  {journal} {\bibinfo
  {journal} {MRS Bull.}\ }\textbf {\bibinfo {volume} {33}},\ \bibinfo {pages}
  {1027} (\bibinfo {year} {2008})}\BibitemShut {NoStop}%
\bibitem [{\citenamefont {Gunkel}\ \emph {et~al.}(2020)\citenamefont {Gunkel},
  \citenamefont {Christensen}, \citenamefont {Chen},\ and\ \citenamefont
  {Pryds}}]{gunkel2020}%
  \BibitemOpen
  \bibfield  {author} {\bibinfo {author} {\bibfnamefont {F.}~\bibnamefont
  {Gunkel}}, \bibinfo {author} {\bibfnamefont {D.~V.}\ \bibnamefont
  {Christensen}}, \bibinfo {author} {\bibfnamefont {Y.~Z.}\ \bibnamefont
  {Chen}},\ and\ \bibinfo {author} {\bibfnamefont {N.}~\bibnamefont {Pryds}},\
  }\href {https://doi.org/10.1063/1.5143309} {\bibfield  {journal} {\bibinfo
  {journal} {Appl. Phys. Lett.}\ }\textbf {\bibinfo {volume} {116}},\ \bibinfo
  {pages} {120505} (\bibinfo {year} {2020})}\BibitemShut {NoStop}%
\bibitem [{\citenamefont {Buchner}\ \emph {et~al.}(2018)\citenamefont
  {Buchner}, \citenamefont {Henne}, \citenamefont {Ney}, \citenamefont
  {Lumetzberger}, \citenamefont {Wilhelm}, \citenamefont {Rogalev},
  \citenamefont {Hen},\ and\ \citenamefont {Ney}}]{buchner2018}%
  \BibitemOpen
  \bibfield  {author} {\bibinfo {author} {\bibfnamefont {M.}~\bibnamefont
  {Buchner}}, \bibinfo {author} {\bibfnamefont {B.}~\bibnamefont {Henne}},
  \bibinfo {author} {\bibfnamefont {V.}~\bibnamefont {Ney}}, \bibinfo {author}
  {\bibfnamefont {J.}~\bibnamefont {Lumetzberger}}, \bibinfo {author}
  {\bibfnamefont {F.}~\bibnamefont {Wilhelm}}, \bibinfo {author} {\bibfnamefont
  {A.}~\bibnamefont {Rogalev}}, \bibinfo {author} {\bibfnamefont
  {A.}~\bibnamefont {Hen}},\ and\ \bibinfo {author} {\bibfnamefont
  {A.}~\bibnamefont {Ney}},\ }\href {https://doi.org/10.1063/1.5023898}
  {\bibfield  {journal} {\bibinfo  {journal} {Journal of Applied Physics}\
  }\textbf {\bibinfo {volume} {123}},\ \bibinfo {pages} {203905} (\bibinfo
  {year} {2018})}\BibitemShut {NoStop}%
\end{thebibliography}%

\end{document}